\documentclass[12pt]{iopart}

\usepackage{iopams}  
\usepackage{graphicx}
\usepackage{adjustbox}
\usepackage{xcolor}
\newtheorem{proposition}{Proposition}
\newtheorem{proof}{Proof}

\newtheorem{relation}{Relation}

\usepackage{eqnarray}
\usepackage{yhmath}
\usepackage{subcaption}

\usepackage{float}
\begin{document}

\title[Relations between ER, EP and IG in LSS]{Relations between entropy rate, entropy production and information geometry in linear stochastic systems}

\author{Adrian-Josue Guel-Cortez$^1$$^2$,Eun-jin Kim$^1$}

\address{$^1$Centre for fluid and complex systems, Coventry University, Coventry, UK.}
\ead{$^2$adrianjguelc@gmail.com}
\vspace{10pt}
\begin{indented}
\item[]August 2017
\end{indented}

\begin{abstract}
In this work, we investigate the relation between the concept of ``information rate'', an information geometric method for measuring the speed of the time evolution of the statistical states of a stochastic process, and stochastic thermodynamics quantities like entropy rate and entropy production. Then, we propose the application of entropy rate and entropy production to different practical applications such as abrupt event detection, correlation analysis, and control engineering. Specifically, by utilising the Fokker-Planck equation of multi-variable linear stochastic processes described by Langevin equations, we calculate the exact value for information rate, entropy rate, and entropy production and derive various inequalities among them. Inspired by classical correlation coefficients and control techniques, we create entropic-informed correlation coefficients as abrupt event detection methods and information geometric cost functions as optimal thermodynamic control policies, respectively. The methods are analysed via the numerical simulations of common prototypical systems.
\end{abstract}

%
\vspace{2pc}
\submitto{\JSTAT}
%
%
%

\section{Introduction}\label{sec:intro}
Information geometry or the application of differential geometry to the information sciences has brought to light new tools for the analysis of dynamical systems \cite{nielsen2020elementary,amari2016information}. For instance, the information length (IL) \cite{kim2021information,kim2021informationb}, given by the time integral of information rate, describes the total amount of statistical changes that a time-varying probability distribution takes through in time. IL appears as an important tool for the analysis of the non-equilibrium processes by providing a possible link between stochastic processes, complexity and geometry. In fact, IL has already been applied with success to different scenarios such as the quantification of hysteresis in forward-backward processes \cite{kim2017geometric,hollerbach2020time}, correlation and self-regulation among different players \cite{hollerbach2020time}, phase transitions \cite{kim2020information}, and prediction of sudden events \cite{guel2021information}. Yet, the thermodynamic significance of IL and information rate seem to be less understood.

To find a connection between IL and thermodynamics, we can use stochastic thermodynamics (ST) \cite{seifert2012stochastic,peliti2021stochastic}. As it has already been popularised, ST makes use of stochastic calculus to draw a correspondence between micro/mesoscopic stochastic dynamics and macroscopic thermodynamics. In other words, we use ST to describe the interaction of a micro/mesoscopic system with one or multiple reservoirs. For instance, the dynamics of a Brownian particle suspended in a fluid in thermodynamic equilibrium is described by a Langevin/Fokker Planck equation. ST introduces time information uncertainty relations in thermodynamics \cite{nicholson2020time}, information theory to causality, modelling and control \cite{lozano2021information}, optimal protocols \cite{Deffner2020}, or fluctuation relations \cite{jarzynski2020fluctuation,seifert2012stochastic}. In addition, it has inspired works in neuroscience \cite{friston2021stochastic}, system dynamics \cite{haddad2019dynamical} and control theory \cite{bechhoefer2021control}.

The main aim of this paper is to elucidate the link between thermodynamic quantities (e.g., entropy rate
\cite{tome2006entropy,cocconi2020entropy}) and information rate (detailed in \S IV) while exploring their application to abrupt event detection, correlation analysis, and control engineering. Even though some relations between information geometry and stochastic thermodynamics have been derived already (for instance, see \cite{Ito2020,Nicholson2018,Ito2018}), this work focuses on generalising a relation initially proposed in \cite{kim2021information,kim2021informationb} for the Ornstein–Uhlenbeck (OU) process governed by a first-order linear stochastic differential equation. Note that, in comparison to \cite{Ito2018} which determines inequalities between information geometry and thermodynamic observables. The relation reported in \cite{kim2021information} connects information geometry to the ``entropy balance equation'' by explicitly using the values of entropy rate and entropy production. We generalise this relation to a set of linear Langevin equations using the corresponding Fokker-Planck equation \cite{guel2020information} for multivariable linear processes. By utilising the exact time-varying probability density function (PDF), we calculate the values of the entropy rate, entropy production rate, and information rate. Then, we establish an inequality and an equality between information rate, entropy production and entropy rate for general linear stochastic systems and fully decoupled systems, respectively. Our results suggest that a weighted value of entropy production plus the square of the entropy rate is generally an upper bound of the information rate and that such a relation is fully determined by the structure of the system's harmonic potential. 

Furthermore, after establishing an information-thermodynamic relation, we propose normalised correlation coefficients based on information rate and entropy production  \cite{duncan1970calculation}, and compare them in the analysis of abrupt events. Specifically, we perform the numerical simulation of a second-order stochastic process where the abrupt events are stimuli in the form of impulse-like functions applied to the system's mean value and covariance matrix. The results suggest that the information rate coefficient performs better when detecting/predicting abrupt events. Moreover, we conduct abrupt events analysis for systems with more than two random variables via the Euclidean norm of the information rate and entropy production as approximate correlation coefficients. Finally, we explore the design of optimal ``static'' feedback control algorithms and the corresponding effects on entropy rate and entropy production for the minimisation of the number of statistical fluctuations via the solution of cost functions using IL and entropy production. 

The rest of the paper is organised as follows. Section \ref{sec:prel} describes the mathematical model and preliminary results to be used throughout the paper. Section \ref{sec:ER} details the calculation of the entropy rate, entropy production and entropy flow in multivariable stochastic systems. Section \ref{sec:ERandIL} introduces the concept of information rate and IL and presents their relation to the thermodynamic quantities of entropy production and entropy rate. In Section \ref{sec:CC}, new normalised correlation coefficients in terms of entropy rate and entropy production are introduced. Section \ref{sec:app} demonstrates the application of the main results through the numerical simulation of different toy models. Section \ref{sec:concl} states the conclusions of the paper and future work. Finally, from \ref{sec:ERProof} to \ref{sec:EquProof}, we present the detailed steps for the derivation of the main results in the paper.

\section{Model}\label{sec:prel}
Consider the following set of Langevin equations
\begin{equation}
    \frac{d x_i}{dt}=f_i(\mathbf{x};t)+\xi_i(t).\label{eqmain}
\end{equation}
Here, $f_i:\mathbb{R}^n\rightarrow \mathbb{R}$ is a function that maps the variables of the vector $\mathbf{x}\in\mathbb{R}^n:=[x_1,x_2,\dots,x_n]^\top$ to a real value at a given time $t\in\mathbb{R}$; $\xi_i$ is a short-correlated random noise satisfying $\langle \xi_i(t)\rangle=0$ and $\langle\xi_i(t)\xi_j(t^\prime)\rangle=2D_{ij}\delta(t-t^\prime)$, where $D_{ij}$ is the amplitude of the correlation function $\langle\xi_i(t)\xi_j(t^\prime)\rangle$. In this work, $f_i(\mathbf{x};t)$ is a linear function defined as
\begin{equation}
f_i(\mathbf{x};t):=\mathbf{A}_i\mathbf{x}(t)+\mathbf{B}_i\mathbf{u}(t)=\sum_{j=1}^{n}a_{ij}x_j(t)+\sum_j^p b_{ij}u_j(t), \label{eq:harmonicp}
\end{equation}
where it becomes clear that $\mathbf{u}(t)\in\mathbb{R}^{p}$ is a vector of continuous time dependant functions such that $\mathbf{u}(t):=[u_1(t),u_2(t),\dots,u_p(t)]^\top$; $\mathbf{A}_i$ and $\mathbf{B}_i$ are the $i$-th row vectors of the constant matrices $\mathbf{A}\in\mathbb{R}^{n\times n}$ and $\mathbf{B}\in\mathbb{R}^{n\times p}$ used to couple the different system states $x_j(t)$ and deterministic forces $u_j(t)$, respectively. Hence, we consider that the set of particles described by \eqref{eqmain} are driven by a harmonic potential ($\mathbf{A}_i\mathbf{x}(t)$), deterministic force ($\mathbf{B}_i\mathbf{u}(t)$ due to $x$-independent but possibly time-dependent $u(t)$), and a stochastic forcing term $\xi_i(t)$. Note that we have used a state space representation \cite{Chen2013} of the term $f_i$ as we are concerned with applications of stochastic thermodynamics/information geometry to dynamical control systems, causality and abrupt events analysis.

The Fokker-Planck equation corresponding to the Langevin equation \eqref{eqmain} is given by
\begin{equation}
    \frac{\partial p(\mathbf{x};t)}{\partial t}\!=\!-\sum_i \frac{\partial}{\partial x_i}\left(f_i(\mathbf{x};t)p(\mathbf{x};t)\right)+\sum_{i}\sum_{j}D_{ij}\frac{\partial^2 p(\mathbf{x};t)}{\partial x_i\partial x_j}. \label{mainFPE}
\end{equation}
Equation \eqref{mainFPE} describes the time evolution of the PDF $p(\mathbf{x};t):\mathbb{R}^n\rightarrow \mathbb{R}$ for $\mathbf{x}:=\{x_1,x_2,\dots,x_n\}$ in \eqref{eqmain}. 
Equation \eqref{mainFPE} can be written in terms of a probability current as follows 
\begin{equation}
 \frac{\partial p(\mathbf{x};t)}{\partial t}=-\sum_i\frac{\partial}{\partial x_i}J_i(\mathbf{x};t),   \label{mainJFPE}
\end{equation}
where $J_i$ is the i-th component of the probability current $J$ defined by
\begin{equation}
J_i(\mathbf{x};t)=f_i(\mathbf{x};t)p(\mathbf{x};t)-\sum_jD_{ij}\frac{\partial}{\partial x_j}p(\mathbf{x};t). \label{maincurrent}
\end{equation}

In general, the solutions of Equation \eqref{mainJFPE} require numerical simulations. However, thanks to the linearity of $f_i(\mathbf{x};t)$ in \eqref{eqmain}, we have an analytical time-dependent solution of Equation \eqref{mainJFPE} given by the following Proposition.
\begin{proposition}\label{prop:GP}
Given an initial multivariable Gaussian PDF $p(\mathbf{x};t_0)$, the solution of $p(\mathbf{x};t)$ in \eqref{mainFPE} is Gaussian and described by \cite{tome2015stochastic}
\begin{equation}
	p(\mathbf{x};t)=\frac{1}{\sqrt{\det(2\pi\boldsymbol{\Sigma})}}e^{Q(\mathbf{x};t)}, \label{mainpdf}
\end{equation}
where $Q(\mathbf{x};t)=-\frac{1}{2}\left(\mathbf{x}-\boldsymbol{\mu}(t)\right)^\top \boldsymbol{\Sigma}^{-1}(t)\left(\mathbf{x}-\boldsymbol{\mu}(t)\right)$; $\boldsymbol{\mu}(t)\in\mathbb{R}^n$ and $\boldsymbol{\Sigma}(t)\in\mathbb{R}^{n\times n}$ are the mean and covariance value of the random variable $\mathbf{x}$.
\end{proposition}
According to Proposition \ref{prop:GP}, the dynamics of the PDF at every instant of time are governed solely by the value of $\boldsymbol{\mu}(t)$ and $\boldsymbol{\Sigma}(t)$. In the case of linear stochastic systems, the value of $\boldsymbol{\mu}(t)$ and $\boldsymbol{\Sigma}(t)$ can be obtained by solving the following set of differential equations \cite{maybeck1982stochastic}
\begin{eqnarray}
    \dot{\mu}_i(t)&=&\sum_j^n a_{ij}\mu_j(t)+\sum_j^p b_{ij}u_j(t),\label{eqmdynamics}\\
    \dot{\Sigma}_{ij}(t)&=&\sum_k^n a_{ik}\Sigma_{kj}(t)+\sum_k^n a_{jk}\Sigma_{ki}(t)+2D_{ij}(t). \label{eqsdynamics}
\end{eqnarray}
Equations \eqref{eqmdynamics} and \eqref{eqsdynamics} can be rewritten in the following matricial form
\begin{eqnarray}
    \dot{\boldsymbol{\mu}}(t)&=&\mathbf{A}\boldsymbol{\mu}(t)+\mathbf{B}\mathbf{u}(t),\label{eqmdynamics2}\\
    \dot{\boldsymbol{\Sigma}}(t)&=&\mathbf{A}\boldsymbol{\Sigma}(t)+\boldsymbol{\Sigma}(t)\mathbf{A}^\top+2\mathbf{D}(t). \label{eqsdynamics2}
\end{eqnarray}
In Equation \eqref{eqsdynamics2}, $\mathbf{D}$ is an square matrix whose elements are $D_{ij}(t)$.

\section{Entropy rate}\label{sec:ER}
Given a time-varying multivariable PDF $p(\mathbf{x};t)$, its entropy rate is defined as \cite{tome2006entropy}
\begin{equation}
    \dot{S}(t)=\frac{d}{dt}S(t)=-\int_{\mathbb{R}^n} \dot{p}(\mathbf{x};t)\text{ln}\left(p(\mathbf{x};t)\right) \rmd^n x.\label{mainER}
\end{equation}
By substituting \eqref{mainFPE} in \eqref{mainER}, we obtain
\begin{equation}
    \frac{d}{dt}S(t)=\int_{\mathbb{R}^n} \left( \sum_i\frac{\partial}{\partial x_i}J_i(\mathbf{x};t)\right)\text{ln}\left(p(\mathbf{x};t)\right)\rmd^n x
    =-\int_{\mathbb{R}^n} \sum_iJ_i(\mathbf{x};t) \left(\frac{\partial}{\partial x_i}\text{ln}\left(p(\mathbf{x};t)\right)\right)\rmd^n x. \label{ERa}
\end{equation}
Now, after substituting \eqref{maincurrent} in \eqref{ERa}, we have
\begin{equation}
    \frac{d}{dt}S(t)=-\int_{\mathbb{R}^n} \sum_i J_i(\mathbf{x};t)\left(\frac{f_i(\mathbf{x};t)}{D_{ii}}-\frac{J_i(\mathbf{x};t)}{D_{ii}p(\mathbf{x};t)}-\frac{\sum_{j\neq i}D_{ij}\frac{\partial}{\partial x_j}p(\mathbf{x};t)}{D_{ii}p(\mathbf{x};t)}\right) \rmd^n x. \label{ERb}
\end{equation}
From \eqref{ERb}, the entropy production rate of the system corresponds to the positive definite part 
\begin{equation}
    \Pi=\int_{\mathbb{R}^n} \sum_i \frac{J_i(\mathbf{x};t)^2}{D_{ii}p(\mathbf{x};t)} \rmd^n x , \label{eprod}
\end{equation}
while the entropy flux (entropy from the system to the environment) is
\begin{equation}
    \Phi\!\!=\!\!\int_{\mathbb{R}^n} \sum_i\left( \frac{J_i(\mathbf{x};t)f_i(\mathbf{x};t)}{D_{ii}}-\frac{\sum_{j\neq i}D_{ij}J_i(\mathbf{x};t)\frac{\partial}{\partial x_j}p(\mathbf{x};t)}{D_{ii}p(\mathbf{x};t)}\right) \rmd^n x.\label{eflux}
\end{equation}
In this paper, we focus on the case when $D_{ij}=0$ if $i\neq j$ to simplify \eqref{eflux} as
\begin{equation}
\Phi=\int_{\mathbb{R}^n} \sum_i\left( \frac{J_i(\mathbf{x};t)f_i(\mathbf{x},t)}{D_{ii}}\right) \rmd^n x. \label{efluxb}
\end{equation}
From Equations \eqref{eprod}-\eqref{efluxb}, we can define the entropy production $\Pi_{J_i}$ and the entropy flow $\Phi_{J_i}$ from the $J_i$ probability current as
\begin{equation} \label{PR:marginals}
\begin{split}
\Pi_{J_i}&=\int_{\mathbb{R}^n}\frac{J_i(\mathbf{x};t)^2}{D_{ii}p(\mathbf{x};t)}\rmd^n x,\\
\Phi_{J_i}&=\int_{\mathbb{R}^n} \frac{J_i(\mathbf{x};t)f_i(\mathbf{x};t)}{D_{ii}}\rmd^n x.    
\end{split}
\end{equation}
Finally, entropy rate can be expressed as follows
\begin{equation}
    \frac{d S}{dt}=\Pi-\Phi, \label{eqEPmain}
\end{equation}
where, from \eqref{PR:marginals}, clearly $\Pi=\sum_i\Pi_{J_i}$ and $\Phi=\sum_i\Phi_{J_i}$. Equation \eqref{eqEPmain} is a well known expression for irreversible processes \cite{landi2013entropy}. 
Notice that \eqref{eprod}-\eqref{eflux} require that $D_{ii}>0$. If $D_{ii}=0$, we have $\Pi_{J_i}=0$ and
\begin{equation}
    \Phi_{J_i}=\left\langle\frac{\partial f_i(\mathbf{x},t)}{\partial x_i}\right\rangle=a_{ii}.
\end{equation}
\subsection{Entropy rate for Gaussian dynamics}
By using $p(\mathbf{x};t)$ given in equation \eqref{mainpdf}, we derive the expressions for entropy production \eqref{eprod} and entropy flux \eqref{efluxb} in terms of $\boldsymbol{\mu},\boldsymbol{\Sigma},\mathbf{A}$ and $\mathbf{D}$ as follows. 
\begin{relation}\label{EPEFGP}
The value of entropy production $\Pi$ and entropy flow $\Phi$ in a Gaussian process whose mean $\boldsymbol{\mu}$ and covariance $\boldsymbol{\Sigma}$ are governed by equations \eqref{eqmdynamics2}-\eqref{eqsdynamics2} are given by 
\begin{eqnarray}
\Pi&=&\dot{\boldsymbol{\mu}}^\top\mathbf{D}^{-1}\dot{\boldsymbol{\mu}}\!+\!{\normalfont\Tr}\left(\mathbf{A}^\top\mathbf{D}^{-1}\mathbf{A}\boldsymbol{\Sigma}\right)+{\normalfont\Tr}\left(\boldsymbol{\Sigma}^{-1}\mathbf{D}\right)\!+\!2{\normalfont\Tr}(\mathbf{A}), \label{SPeq}\\
\Phi&=&\dot{\boldsymbol{\mu}}^\top\mathbf{D}^{-1}\dot{\boldsymbol{\mu}}\!+\!{\normalfont\Tr}\left(\mathbf{A}^\top\mathbf{D}^{-1}\mathbf{A}\boldsymbol{\Sigma}\right)
+{\normalfont\Tr}(\mathbf{A}). \label{SFeq}
\end{eqnarray}
\end{relation}
\begin{proof}
See \ref{sec:ERProof}.
\end{proof}
From Relation \ref{EPEFGP}, we readily obtain the entropy rate (for further details, see \eqref{TSeqa} in \ref{sec:ERProof})
\begin{equation}
    \frac{dS}{dt}=\frac{1}{2}\frac{d}{dt}\text{ln}|\boldsymbol{\Sigma}|=\frac{1}{2}\Tr\left(\boldsymbol{\Sigma}^{-1}\dot{\boldsymbol{\Sigma}}\right). \label{Erate}
\end{equation}
Clearly, equation \eqref{Erate} can also be obtained directly from the time derivative of the entropy $S$ in a Gaussian Process
\begin{equation}
S = \frac{n}{2} \ln(2\pi) + \frac{1}{2} \ln|\boldsymbol{\Sigma}| + \frac{1}{2} n.    
\end{equation}
Again, if $D_{ii}=0$ $\forall i=1,2,\dots,n$, we have $\Pi=0$ and
\begin{equation}
    \Phi=\Tr(\mathbf{A}).
\end{equation}
Expressions similar to \eqref{SPeq}-\eqref{Erate} but with an emphasis in distinguishing between variables that are even and odd under time reversal can be found in \cite{landi2013entropy}.
\section{Relation between Entropy rate and information rate} \label{sec:ERandIL}
As noted previously in Section \ref{sec:intro}, we aim to elucidate the link between entropy production $\Pi$, entropy rate $\dot{S}$, and information rate. We recall that for a time-varying multivariable PDF $p(\mathbf{x};t)$, we define its IL $\mathcal{L}$ as
\begin{equation}
\mathcal{L}(t)=\int_{0}^{t}\left(\sqrt{\int_{\mathbb{R}^n} p(\mathbf{x};\tau)\left[\partial_\tau\ln{p(\mathbf{x};\tau)}\right]^2\rmd^n x}\right)\rmd \tau
=\int_{0}^{t}\Gamma(\tau) \rmd \tau, \label{defIL}
\end{equation}
where $\Gamma(\tau) = \sqrt{\int_{\mathbb{R}^n} p(\mathbf{x};\tau)\left[\partial_\tau\ln{p(\mathbf{x};\tau)}\right]^2\rmd^n x}$ is called the information rate and $\Gamma^2(\tau)$ the information energy. The value of $\Gamma^2(\tau)$ can also be understood as the Fisher information where the time is the control parameter \cite{kim2021informationb}. Since $\Gamma$ gives the rate of change of $p(\mathbf{x};t)$, its time integral $\mathcal{L}$ quantifies the amount of statistical changes that the system goes through in time from the initial PDF $p(\mathbf{x};0)$ to a final PDF $p(\mathbf{x};t)$ \cite{guel2020information}.  

When $p(\mathbf{x};t)$ is a Gaussian PDF, the information rate $\Gamma$ of the joint PDF takes the compact form \cite{kim2021information,guel2020information,malago2015}
\begin{equation}
    \Gamma^2=\dot{\boldsymbol{\mu}}^\top\boldsymbol{\Sigma}^{-1}\dot{\boldsymbol{\mu}}+\frac{1}{2}\Tr\left((\boldsymbol{\Sigma}^{-1}\dot{\boldsymbol{\Sigma}})^2\right).\label{defirate}
\end{equation}
As it is mentioned in \cite{kim2021information}, if the PDF of Equation \eqref{eqmain} is described by an univariate Gaussian PDF (i.e. the Ornstein–Uhlenbeck (OU) process) the information rate $\Gamma$ is related to the entropy rate $\dot{S}$ and the entropy production $\Pi$ via
\begin{equation}
    \Gamma^2=\frac{D}{\Sigma}\Pi +\dot{S}^2. \label{equalityERI}
\end{equation}
Equation \eqref{equalityERI} can easily be confirmed after some algebra with the following expressions
\begin{eqnarray}
\Pi&=&\frac{\dot{\mu}^2}{D}+\frac{a^2\Sigma}{D}+\frac{D}{\Sigma}+2a=\frac{\dot{\mu}^2}{D}+\frac{\dot{\Sigma}^2}{4\Sigma D},\nonumber\\
\dot{S}&=&\frac{1}{2}\frac{\dot{\Sigma}}{\Sigma},\quad
\Gamma^2=\frac{\dot{\mu}^2}{\Sigma}+\frac{1}{2}\left(\frac{\dot{\Sigma}}{\Sigma}\right)^2, \label{eqsdemo}
\end{eqnarray}
where $\mu,\Sigma,D$ and $a$ are the scalar version of $\boldsymbol{\mu},\boldsymbol{\Sigma},\mathbf{D}$ and $\mathbf{A}$, respectively, in Proposition \ref{EPEFGP}. 
Extending \eqref{equalityERI} to the case of a $n$-th order Gaussian process is one of the main contributions of this work. Such a result is given by the following relations.
\begin{relation}\label{mainth}
Given an $n$-th order Gaussian process whose mean and covariance are described by Equations \eqref{eqmdynamics2}-\eqref{eqsdynamics2}, a relationship between entropy production $\Pi$, entropy rate $\dot{S}$, and information rate $\Gamma$ is given by
\begin{equation}
    0\leq\Gamma^2\leq\mathcal{E}_u:=\Tr(\boldsymbol{\Sigma}^{-1})\Pi\Tr(\mathbf{D})+\dot{S}^2-2g(\mathbf{s}), \label{mainresult}
\end{equation}
where $\mathbf{s}=[\dot{S}_{J_1},\dot{S}_{J_2},\dots,\dot{S}_{J_n}]^\top$, $g(\mathbf{s}):=\sum_{i<j}^n \dot{S}_{J_i}\dot{S}_{J_j}$ and $\dot{S}_{J_i}$ is the contribution to entropy rate by the current flow $J_i$, i.e.
\begin{equation}
    \dot{S}_{J_i}=-\int_{\mathbb{R}^n} \frac{\partial}{\partial x_i}J_i(\mathbf{x};t)\ln{(p(\mathbf{x};t))} \rmd^n x=\Pi_{J_i}-\Phi_{J_i}.\label{mardS}
\end{equation}
\end{relation}
\begin{proof}
See \ref{sec:appproofmainth}.
\end{proof}
Relation \ref{mainth} provides an inequality between information rate $\Gamma$ (an information metric), entropy rate $\dot{S}$ and entropy production $\Pi$ where the entropy rate $\dot{S}_{J_i}$ of each current flow $J_i$ is explicitly taken into account. In the relation, $\Pi$ is normalised by the product between the amplitude of the environmental fluctuations and the inverse of the system dynamics covariance due to its nature as an extensive quantity \cite{kim2021information}. If $\boldsymbol{\Sigma}$ is constant, the normalised entropy production dominates the square of the information rate. On the other hand, when the system fluctuations change, the flow of information between the system and its environment is considered by the term $\dot{S}^2-2g(\mathbf{s})$ in \eqref{mainresult}. In addition, from Relation \ref{mainth} we have 
\begin{equation}
    \mathcal{L}(t)\leq\mathcal{L}_u(t):= \int_{0}^t\sqrt{\mathcal{E}_u(\tau)}\rmd\tau. \label{eq:upIL}
\end{equation}
Since minimising $\mathcal{L}_u$ will minimise $\mathcal{L}$, we can obtain both a minimum entropy production and a minimum statistical variability behaviour through $\mathcal{L}_u$. 

For unstable systems, we can avoid the computation of the term $g(\mathbf{s})$ involving the contribution to entropy rate by each current flow $J_i$ by using the following relation.
\begin{relation}\label{mainth2}
Given the same conditions as in Relation \ref{mainth}, but considering that the eigenvalues $\varphi_i\in \mathbb{C}$ of the matrix $\mathbf{A}$ satisfy the following inequality
\begin{equation}
    \Re\{\varphi_i\}> 0\quad \forall i=1,2,\dots,n.
\end{equation}
Then, the following result holds
\begin{equation}
    0\leq\Gamma^2\leq\mathcal{I}_u:= \frac{1}{n}\normalfont{\Tr}(\boldsymbol{\Sigma}^{-1})\Pi_u\normalfont{\Tr}(\mathbf{D})+\dot{S}^2, \label{mainresult2}
\end{equation}
where $\Pi_u\geq\Pi$ (an upper bound of entropy production) defined by
\begin{equation}
\Pi_u:=\normalfont{\Tr}(\dot{\boldsymbol{\mu}}\dot{\boldsymbol{\mu}}^\top)\normalfont{\Tr}(\mathbf{D}^{-1})\!+\!\frac{1}{4}\normalfont{\Tr}(\boldsymbol{\Sigma}^{-1})\normalfont{\Tr}(\dot{\boldsymbol{\Sigma}})^2\normalfont{\Tr}(\mathbf{D}^{-1}). \label{upbmainresult}
\end{equation}
\end{relation}
\begin{proof}
See \ref{sec:IneProof}.
\end{proof}
Now, we investigate the case when a relation between $\Gamma$, $\dot{S}$ and $\Pi$ can be expressed in the form of equality. If and only if $\mathbf{A}$ in \eqref{eqmdynamics2}-\eqref{eqsdynamics2} is a diagonal matrix, i.e. we have a set of linearly independent stochastic differential equations (this can be after applying decoupling transformations \cite{Falb1967}), the following result holds. 
\begin{relation}\label{th3}
Given a $n$-th order Gaussian process where all its random variables are independent, we have 
\begin{equation}
    \Gamma^2:=\sum_i\frac{D_{ii}}{\Sigma_{ii}}\Pi_i+\sum_i\dot{S}_i^2, \label{eqcol1}
\end{equation}
where $\Pi_i$ and $\dot{S}_i$ are the entropy production and entropy rate from the marginal PDF $p(x_i,t)$ of $x_i$, respectively.
\end{relation}
\begin{proof}
See \ref{sec:EquProof} .
\end{proof}
Again, the value of $\Pi_i$ in \eqref{eqcol1} is given by Equation \eqref{PR:marginals} while $\dot{S}_i$ is given by \eqref{mardS}.
From \eqref{eqcol1}, it is inferred that a first order system is a special case of Relation \ref{th3}. In addition, Equation \eqref{eqcol1} tells us that the geodesic (length-minimising curve between the initial and final PDF) of $\mathcal{L}(t)$ can be computed utilising the entropy rate and entropy production values (for further details on the geodesic problem, see \cite{kim2016geometric}). More importantly, since Relation \ref{th3} permits us to equate the effects of IL geodesic dynamics on the system stochastic thermodynamics, Equation \eqref{eqcol1} can be used as part of a cost function employed to design controls that lead to both high energetic efficiency and minimum variability system closed-loop responses. (for further details, see \cite{kim2021information,kim2021informationb}). 
\section{On correlation coefficients}\label{sec:CC}
In this section, we define different types of correlations coefficients in terms of entropy production and information rate utilising mutual information and Pearson correlation coefficient, which will be used for abrupt event detection and causality analysis as discussed in Section \ref{sec:app}.
\subsection{Mutual information}
The mutual information between two continuous random variables $x_i$ and $x_j$ with a joint PDF
$p(\mathbf{x};t)$ at time $t$ is defined as
\begin{equation}
    I_{ij}(t):=\int_{\mathbb{R}^2}p(\mathbf{x};t)\ln{\left(\frac{p(\mathbf{x};t)}{p(x_i,t)p(x_j,t)}\right)}\rmd^2 x
    =S_i(t)+S_j(t)-S(t). \label{minformation}
\end{equation}
Here, $p(x_i,t)$ and $p(x_j,t)$ are the marginal PDFs of the random variables $x_i$ and $x_j$, respectively. Hence, the sub-index $i$ in the entropy $S$ refers to the entropy from the marginal PDF of $x_i$ and its value is simply
\begin{equation}
    S_i(t)=\frac{1}{2}+\ln{\left(\sqrt{2\pi\Sigma_{ii}(t)}\right)}.
\end{equation}
Mutual information represents the amount of information of a random variable that can be obtained by observing another random variable. Hence, it is a measure of the mutual dependence between the two variables \cite{duncan1970calculation}.
To measure correlations between two random variables in a process, we can utilise common normalised variants of the mutual information, for instance, the total correlation formula \cite{nguyen2018entropy,cahill2010normalized}
\begin{equation}
    \rho_I(t):=2\frac{I_{ij}(t)}{S_i(t)+S_j(t)}, 
    \label{Tcorrelation}
\end{equation}
where $x_i$ and $x_j$ are treated symmetrically. Equation \eqref{Tcorrelation} is a weighted average of the asymmetrical uncertainty coefficients $\mathcal{C}_{ij}(t)$ and $\mathcal{C}_{ji}(t)$, defined as
\begin{equation}
    \mathcal{C}_{ij}(t):=\frac{I_{ij}(t)}{S_i(t)}, \quad     \mathcal{C}_{ji}(t):=\frac{I_{ij}(t)}{S_j(t)}, \label{eq_UC}
\end{equation}
weighted by the entropy of each variable separately \cite{press2007numerical}. The uncertainty coefficient \eqref{eq_UC} gives a value between $0$ and $1$, indicating no association or complete predictability of $x_i$ from $x_j$ (given $x_j$, what fraction of $x_i$ we can predict), respectively. Thus, \eqref{Tcorrelation} gives an average of the predictability between $x_i$ and $x_j$. The total correlation formula \eqref{Tcorrelation} is as an alternative to the well-known Pearson correlation coefficient 
\begin{equation}
    \rho:=\frac{\Sigma_{ij}}{\sqrt{\Sigma_{ii}\Sigma_{jj}}}, \label{pCC}
\end{equation}
when dealing with non-linear relationships between the random variables \cite{veyrat2009mutual,li1990mutual,dionisio2004mutual}. Pearson correlation coefficient $\rho$ and mutual information $I$ are related to each other through the following expression 
\begin{equation}
    I_{ij}=-\frac{1}{2}\log\left(1-\rho^2\right).
\end{equation}

\subsection{Information rate and entropy production correlation coefficients}
\begin{figure}[t]
    \centering
    \includegraphics[trim={12cm 5cm 17cm 3.5cm},clip,width=0.4\columnwidth]{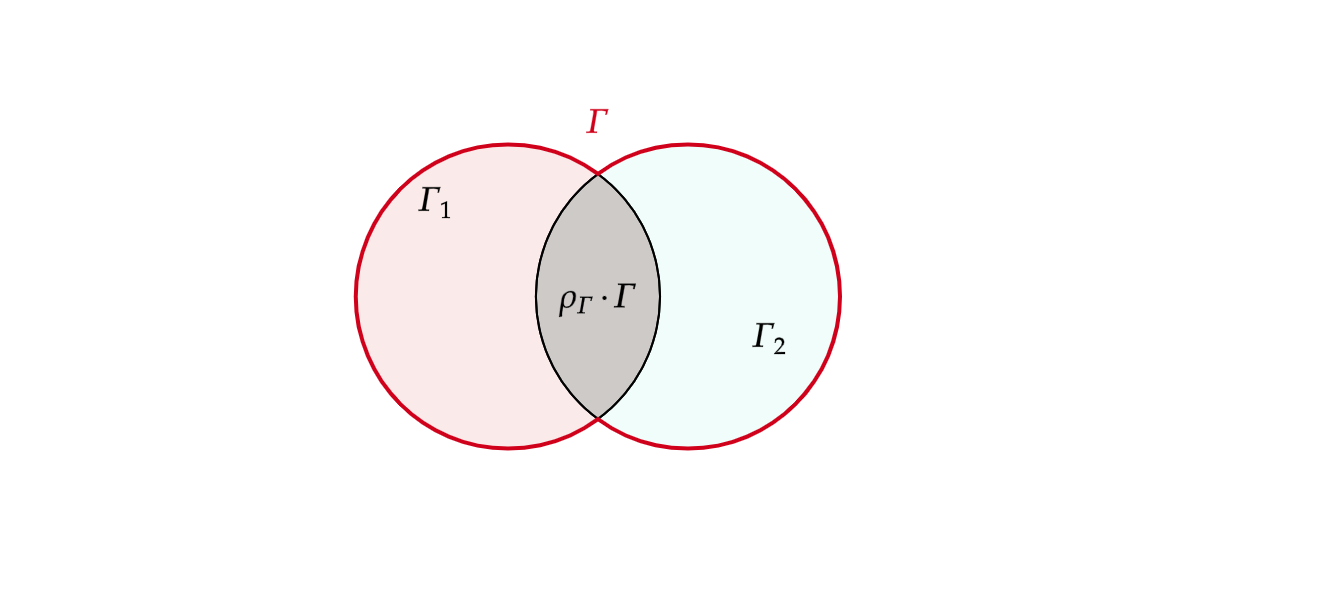}
    \caption{Venn diagram describing the meaning of \eqref{Ecorr}. A similar diagram can be made for \eqref{Picorr}.}
    \label{fig:Venn}
\end{figure}

In analogy to \eqref{Tcorrelation} to \eqref{pCC}, we define new normalised correlation coefficients between two variables $x_i$ and $x_j$ in terms of information rate and entropy production as follows
\begin{eqnarray}
    \rho_\Gamma(t)&:=&\frac{\Gamma_i(t)+\Gamma_j(t)-\Gamma(t)}{\Gamma(t)}, \label{Ecorr}\\
    \rho_\Pi(t)&:=&\frac{\Pi_i(t)+\Pi_j(t)-\Pi(t)}{\Pi(t)}.\label{Picorr}
\end{eqnarray}
Here, $\Pi_i$ and $\Pi_j$ are the contributions from the variable $x_i$ and $x_j$ to the entropy production $\Pi$ (see Equation \eqref{eqindpi}). The values of $\Gamma_i$ and $\Gamma_j$ are the information rates from the marginal PDFs of $x_i$ and $x_j$, respectively. For instance, given the marginal PDF $p(x_i,t)$ of the random variable $x_i$ the value of $\Gamma_i$ is defined as follows
\begin{equation}
\Gamma_i:=\sqrt{\int_{\mathbb{R}} p(x_i;t)\left[\partial_\tau\ln{p(x_i,t)}\right]^2\rmd x_i}.
\end{equation}
Equations \eqref{Ecorr}-\eqref{Picorr} are not defined exactly as the Pearson correlation coefficient \eqref{pCC} or the normalised correlation coefficient of the mutual information \eqref{Tcorrelation}. Instead, they are expressed analogously to the information quality ratio, a quantity of the amount of information of a variable based on another variable against total uncertainty \cite{wijaya2017information}. Hence, $\rho_\Gamma$/$\rho_\Pi$ is said to quantify the predictability of information rate/entropy production of a variable based on another variable. In other words, these coefficients give a weight of how much information one variable can add to the information rate or entropy production of another variable. Also, importantly, these quantities can work for any nonlinear relation. A graphical description of Equation \eqref{Ecorr} in the form of Venn diagram is shown in Figure \ref{fig:Venn}.
\section{Applications}\label{sec:app}
So far, we have introduced a complete set of tools that establish a connection between information geometry and thermodynamics for linear systems. In this section, we apply these to the different linear stochastic toy models and scenarios. Firstly, we start by exploring a simple second order stochastic model \S\ref{sec:sop} representing a harmonically bound particle. Secondly, we will explore a simple decoupled third order stochastic model \S\ref{sec:tdc} commonly used to represent the simplified dynamics of an optical trap in the three-dimensional space. Thirdly, we will investigate higher order models to elucidate the effects of system's dimension on entropy production \S\ref{sec:hos}. Fourthly, we will provide the analysis of abrupt events/perturbations in the dynamics of a second and fourth order stochastic system \S\ref{sec:aea}. Finally, we will explore two optimisation problems for the design of thermodynamic control algorithms \S\ref{sec:mep}.
\subsection{Second and third order process}\label{sec:sop}
\subsubsection{Harmonically bound particle}

Consider the following Langevin equation describing a harmonically bound particle
\begin{equation}
\begin{bmatrix}
\dot{x}_1(t) \\ \dot{x}_2(t)
\end{bmatrix}=\begin{bmatrix}
0 & 1 \\ -\omega^2 & -\gamma
\end{bmatrix}\begin{bmatrix}
x_1(t)\\x_2(t)
\end{bmatrix}+\begin{bmatrix}
\xi_1(t) \\ \xi_2(t)
\end{bmatrix}, \label{bmotiona}
\end{equation}
where the parameters $\omega$ and $\gamma$ are related to the system's natural frequency and damping, respectively. 

\begin{figure}[h]
    \centering
    \begin{subfigure}[b]{0.49\textwidth}
    \centering
    \includegraphics[trim={0.8cm 0cm 1.3cm 0cm},clip,width=\columnwidth]{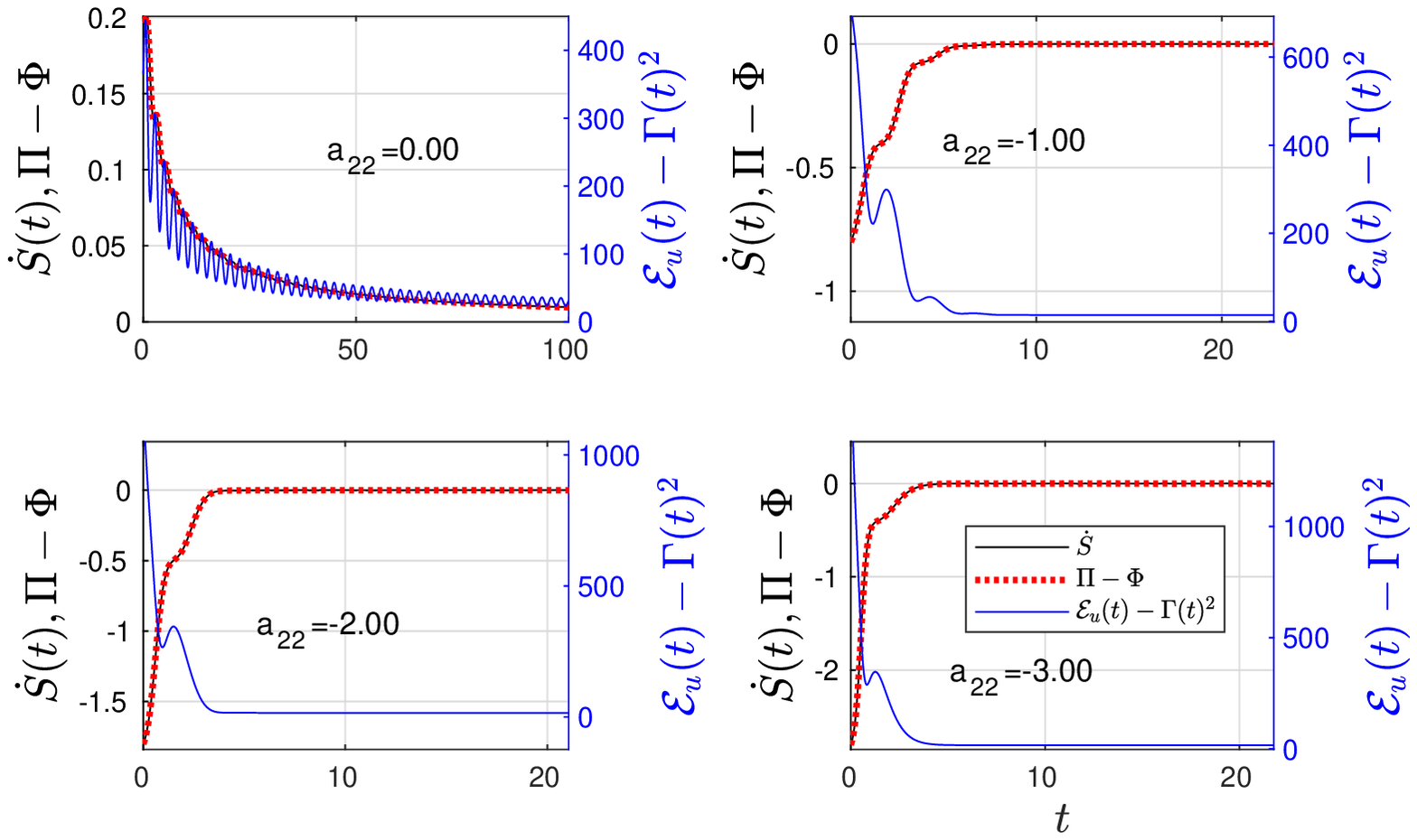}
    \caption{$a_{11}=0,a_{12}=1$, $a_{21}=-2,a_{22}=\{0,-1,-2,-3\}$. }
    \label{fig:ERexp1a}
    \end{subfigure}
    \begin{subfigure}[b]{0.49\textwidth}
    \centering
    \includegraphics[trim={0.8cm 0cm 1.3cm 0cm},clip,width=\columnwidth]{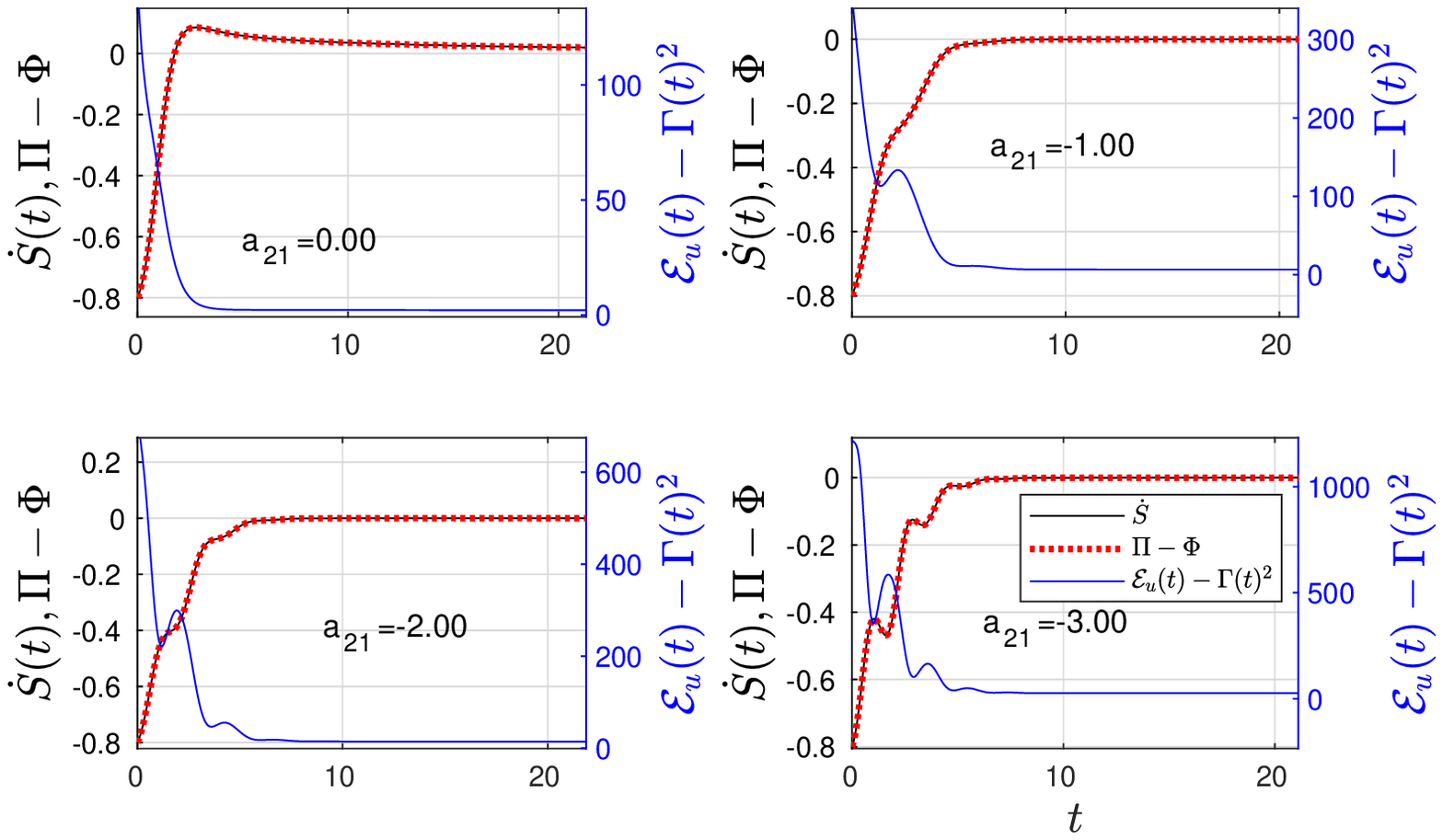}
    \caption{$a_{11}=0,a_{12}=1$, $a_{21}=\{0,-1,-2,-3\},a_{22}=-1$.}
    \label{fig:ERexp1b}
    \end{subfigure}
    \caption{Computational experiment of a second order Langevin equation using $D_{11}=D_{22}=0.01$, $x(0)=1$, $y(0)=1$, $\Sigma_{11}^0=\Sigma_{22}^0=0.1$ and $\Sigma_{12}^0=\Sigma_{21}^0=0$.} \label{fig:ERexp1}
\end{figure}

First, to explore Relations \ref{EPEFGP} and \ref{mainth}, in Figure \ref{fig:ERexp1} we plot the changes on entropy rate $\dot{S}$ computed by using equation \eqref{Erate} and compared them with the value of $\Pi-\Phi$ obtained from equations \eqref{SPeq}-\eqref{SFeq}, confirming the expected relation $\dot{S}=\Pi-\Phi$. Second, to briefly demonstrate that $\mathcal{E}_u>\Gamma^2$, we also show the difference between $\Gamma^2$ (using equation \eqref{defirate}) and $\mathcal{E}_u$ (from Relation \ref{mainth}). Our simulations were done for fixed value of $\omega$ by varying the value of  $\gamma$ (Figure \ref{fig:ERexp1a}) and vice-versa (Figure \ref{fig:ERexp1b}).

As can be concluded from Figure \ref{fig:ERexp1a}, for an undamped harmonic oscillator with $\gamma=0$, the value of $\mathcal{E}_u-\Gamma^2$ tends to decrease with time, meaning that they become equal over time. Here, $\Gamma,\dot{S}>0$ $\forall t\geq 0$ because the system is permanently oscillating. Once we increase $\gamma$, the system goes to the equilibrium giving $\Gamma,\dot{S}\to 0$ and $\mathcal{E}_u\geq 0$ due to $\Pi=\Phi$. In general, for any $\mathbf{A}$ whose eigenvalues have negative real part and $u(t)=0$, entropy production $\Pi$ and entropy flow $\Phi$ in the long-time limit take the following values
\begin{eqnarray}
    \lim_{t\to\infty}\Pi(t)\!\!&=&\!\!\Tr\left(\mathbf{A}^\top\mathbf{D}^{-1}\mathbf{A}\boldsymbol{\Sigma}(\infty)+\boldsymbol{\Sigma}^{-1}(\infty)\mathbf{D}+2\mathbf{A}\right), \nonumber\\
    \lim_{t\to\infty}\Phi(t)\!\!&=&\!\!\Tr\left(\mathbf{A}^\top\mathbf{D}^{-1}\mathbf{A}\boldsymbol{\Sigma}(\infty)+\mathbf{A}\right), \label{limitpi}
\end{eqnarray}
where $\boldsymbol{\Sigma}(\infty)=2\lim_{t\to\infty}\{\int_{0}^{t}e^{\mathbf{A}(t-\tau)}\mathbf{D}e^{\mathbf{A}^\top(t-\tau)}\rmd\tau\}.$ As we see, the transient behaviour and longtime limit of entropy production and entropy flow are fully determined by the value of $e^{\mathbf{A}t}$, which in turn (obviously) depends on the eigenvalues of the matrix $\mathbf{A}$. In \S \ref{sec:mep}, we discuss how such eigenvalues can be modified through a control algorithm (for example, using a full-state feedback control method \cite{Sontag2013}). In system \eqref{bmotiona}, the bigger the value of $\gamma$ the quicker we arrive to equilibrium. On the other hand, increasing the value of $\omega$ with $\gamma>0$ increments the oscillations on the transitory response (see Figures \ref{fig:ERexp1a} and \ref{fig:ERexp1b}) \cite{guel2021information,guel2020information}. 
\subsubsection{Correlation coefficients}
\begin{figure}[h]
    \centering
    \begin{subfigure}[b]{0.7\columnwidth}
    \centering
    \includegraphics[trim={1cm 0cm 1cm 0cm},clip,width=\columnwidth]{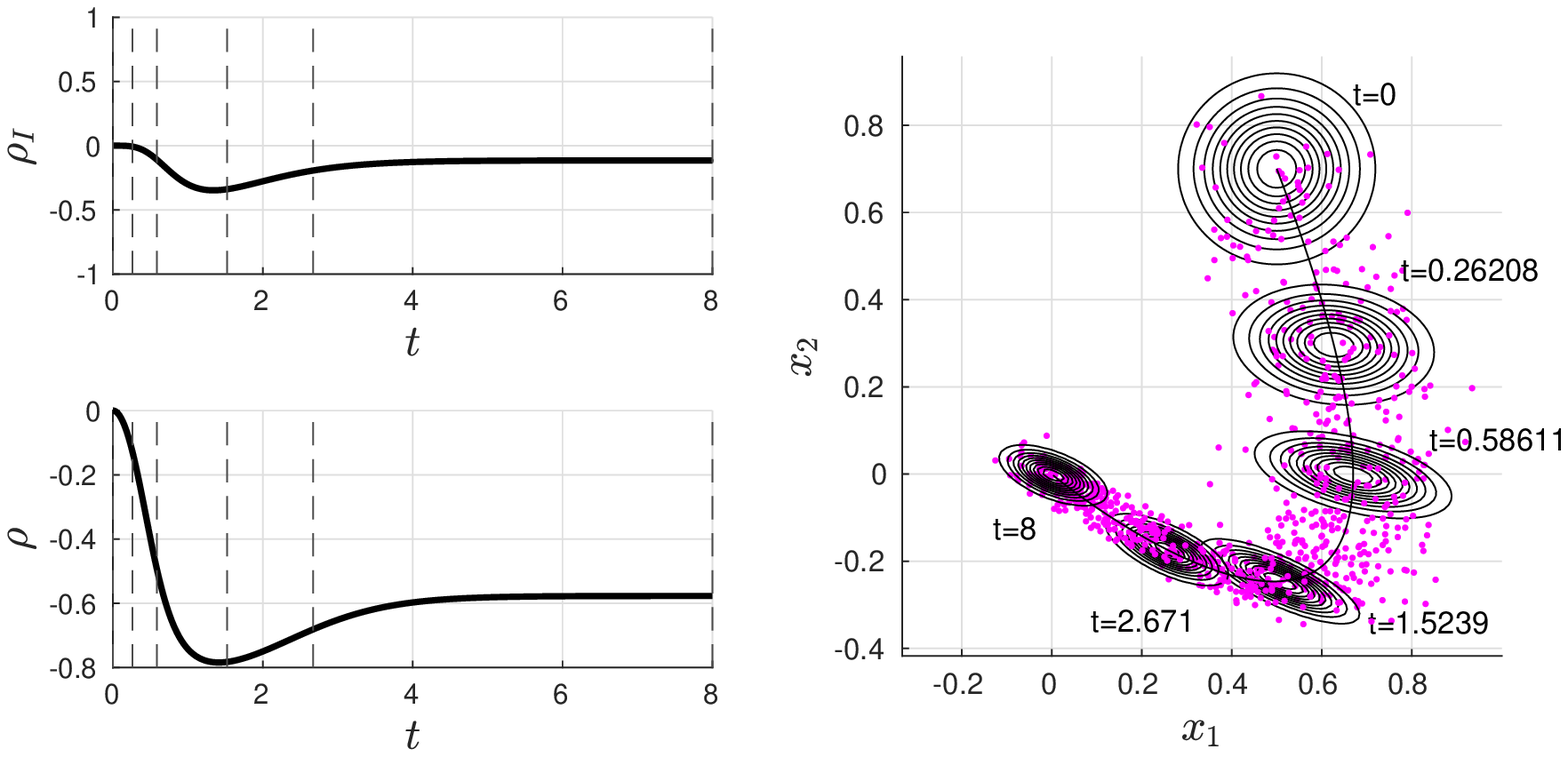}\vspace{-0.7cm}
    \caption{}\label{fig:correlationsa}
    \end{subfigure}
    \begin{subfigure}[b]{0.7\columnwidth}
    \centering
    \includegraphics[trim={1cm 0cm 1cm 0cm},clip,width=\columnwidth]{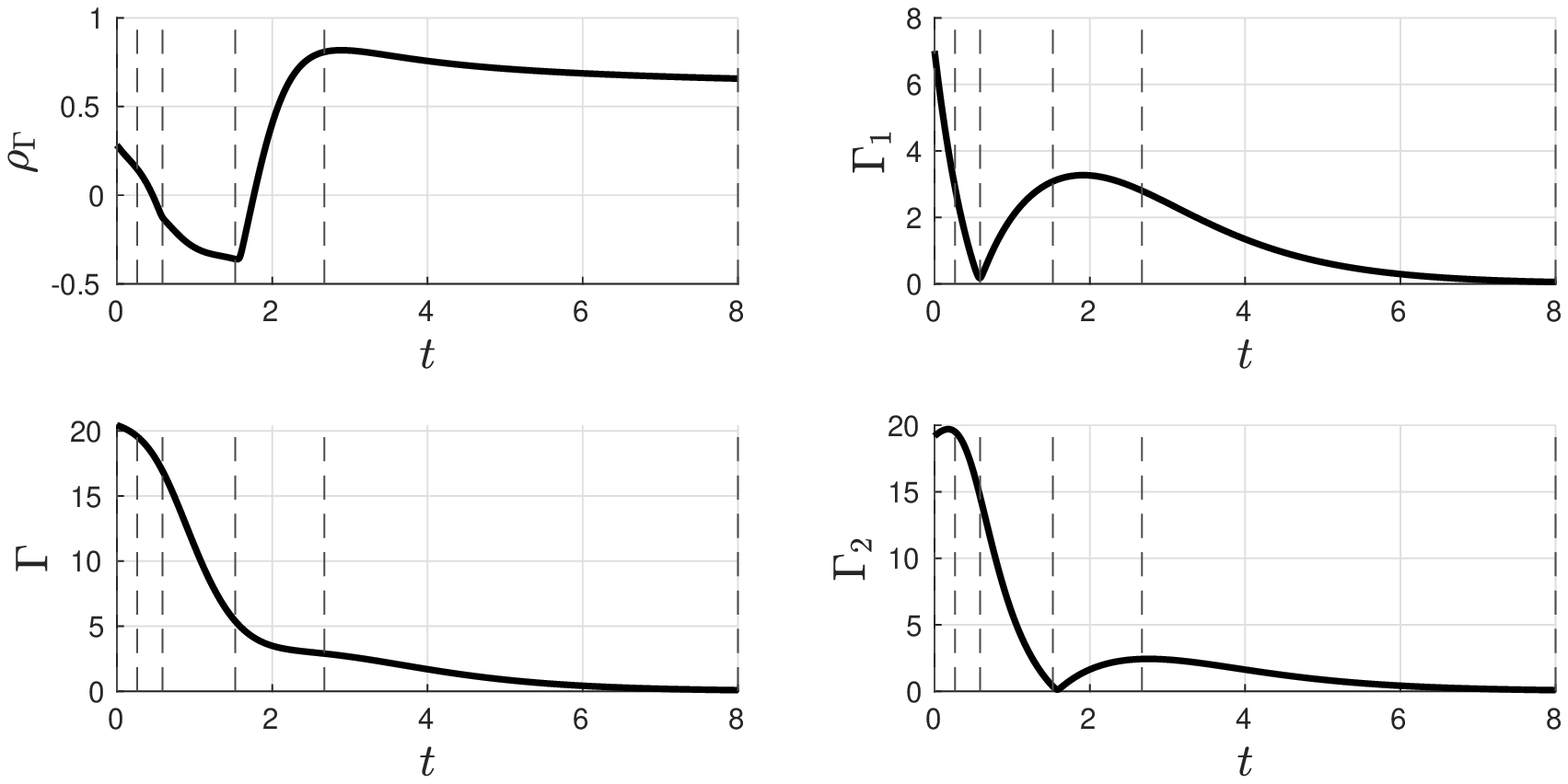}\vspace{-0.7cm}
    \caption{}\label{fig:correlationsb}
    \end{subfigure}
    \begin{subfigure}[b]{0.7\columnwidth}
    \centering
    \includegraphics[trim={0.5cm 0cm 1cm 0cm},clip,width=\columnwidth]{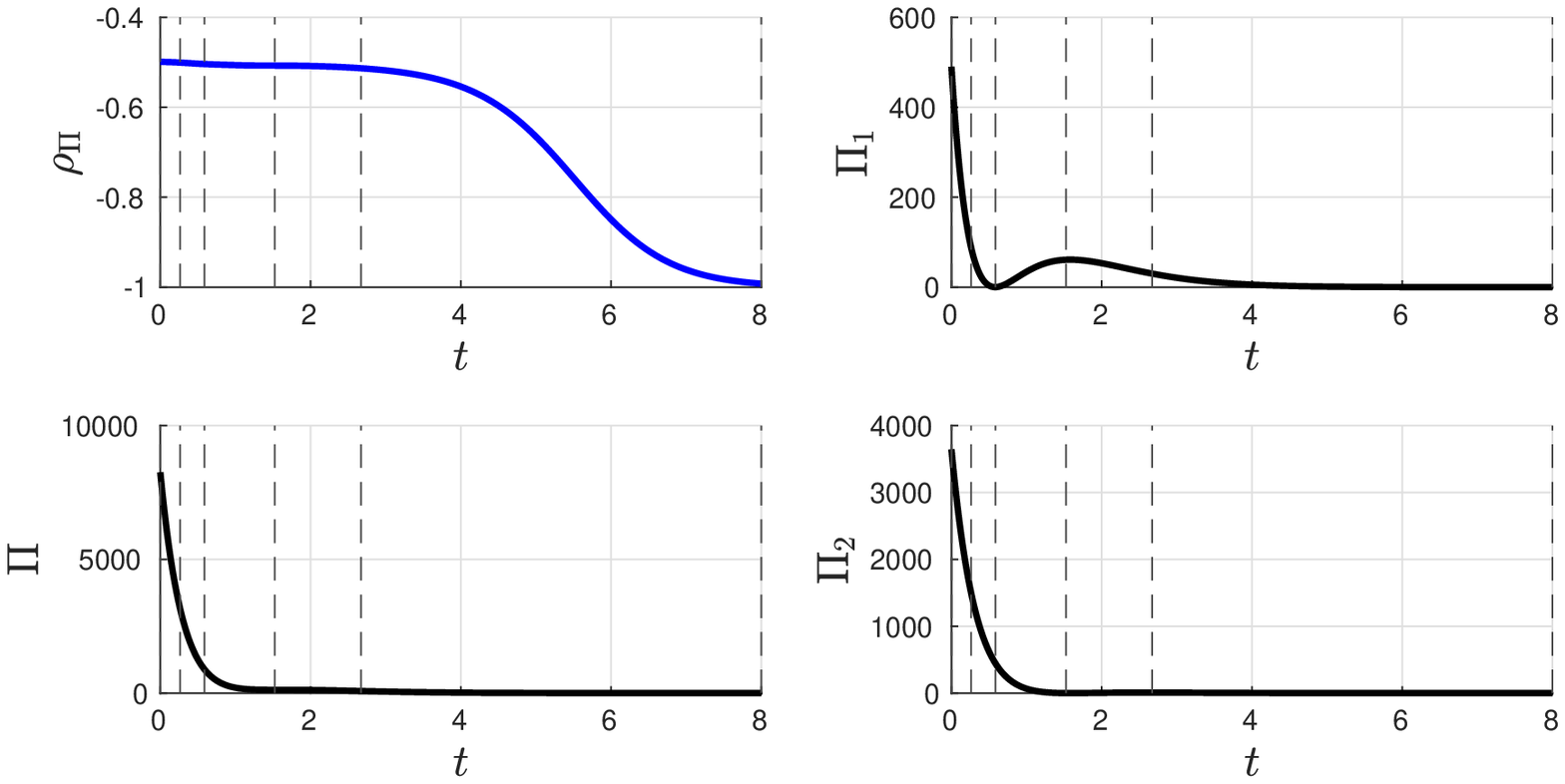}\vspace{-0.7cm}
    \caption{}\label{fig:correlationsc}
    \end{subfigure}    
    \caption{Time evolution of $\rho_\Gamma$, $\rho_{\Pi}$, 
    $\rho_I$, $\rho$, $\Gamma_1$, $\Gamma_2$, $\Pi_1$ and $\Pi_2$ in comparison with the evolution of the snapshots of the phase portraits of $x_1$ vs $x_2$. The parameters used in this simulation are $\omega=1,\gamma=2$, $\mu_1(0)=0.5,\mu_2(0)=0.7$, $\boldsymbol{\Sigma}(0)=0.01\mathbf{I}$ and $\mathbf{D}=0.001\mathbf{I}$ where $\mathbf{I}$ is the identity matrix.}
    \label{fig:correlations}
\end{figure}

Using a bivariate PDF that satisfies \eqref{bmotiona}, we will compare the different correlation coefficients defined in Section \ref{sec:CC}. To this end, Figure \ref{fig:correlations} shows the time evolution of $\rho_\Gamma,\rho_\Pi,\rho_I$ and $\rho$; the phase portraits of $x_1$ vs $x_2$, $\Gamma_1$ vs $\Gamma_2$ and $\Pi_1$ vs $\Pi_2$. The parameters used in this simulation are $\omega=1,\gamma=2$, $\mu_1(0)=0.5,\mu_2(0)=0.7$, $\boldsymbol{\Sigma}(0)=0.01\mathbf{I}$ and $\mathbf{D}=0.001\mathbf{I}$ where $\mathbf{I}$ is the identity matrix. To facilitate the analysis, the phase portrait shows time snapshots with the complete contour of the PDF where times are marked by vertical dashed lines.

The time evolution of the Pearson correlation coefficient $\rho$ and the total correlation coefficient $\rho_I$ may be the easiest coefficients to interpret since their values are related to the shape of the bivariate PDF. For instance, Figure \ref{fig:correlationsa} shows that $\rho$ has negative sign because of the negative slope on the ``linear'' correlation between $x_1$ and $x_2$. Similarly, the total correlation coefficient $\rho_I$ is negative with a maximum of $-0.34$ at some $0.68135<t<1.9323$. Recall that an advantage of the mutual information coefficient $\rho_I$ over the Pearson coefficient $\rho$ is that $\rho_I$ does not assume whether the association of the random variables is linear or not while $\rho$ does. This causes $\rho$ to be zero even when the variables are still stochastically dependent (for further details, see \cite{veyrat2009mutual,li1990mutual,dionisio2004mutual}).

Understanding the meaning of $\rho_\Gamma$ and $\rho_\Pi$ is a bit more complicated as its behaviour cannot be easily related to the shape of the PDF through time. This is because even though the values of $\rho_\Gamma$ and $\rho_\Pi$ clearly take into account the correlation effects through $\boldsymbol{\Sigma}$, they are also sensitive to the mean values of $x_1$ and $x_2$. Hence, they should be interpreted by comparing the time evolution of the pair of quantities whose correlation is studied. 
For instance, Figure \ref{fig:correlationsb} shows that $\rho_\Gamma>0$ in the time intervals where the time evolution of $\Gamma_1$ and $\Gamma_2$ have the same tendency or similar behaviour. $\rho_\Gamma$ reaches its maximum value when $\Gamma_1$ and $\Gamma_2$ are almost similar in amplitude and rate of change. The different behaviour in the time evolution of $\Gamma_1$ and $\Gamma_2$ at $0.26 \lessapprox  t  \lessapprox2$ depicts an anticorrelation which is corroborated by a value of $\rho_\Gamma<0$. Regarding the thermodynamic effects, Figure \ref{fig:correlations} shows that $\Pi_1$ and $\Pi_2$ are always anticorrelated as $\rho_\Pi<0$ at all $t$. The monotonically decrement of $\rho_\Pi$ represents a continously increasing loss of predictability of $\Pi_2$ by $\Pi_1$. 

\subsubsection{Three-dimensional decoupled process}\label{sec:tdc}
\begin{figure}[h]
    \centering
    \includegraphics[trim={10cm 4cm 13cm 0cm},clip,width=0.4\columnwidth]{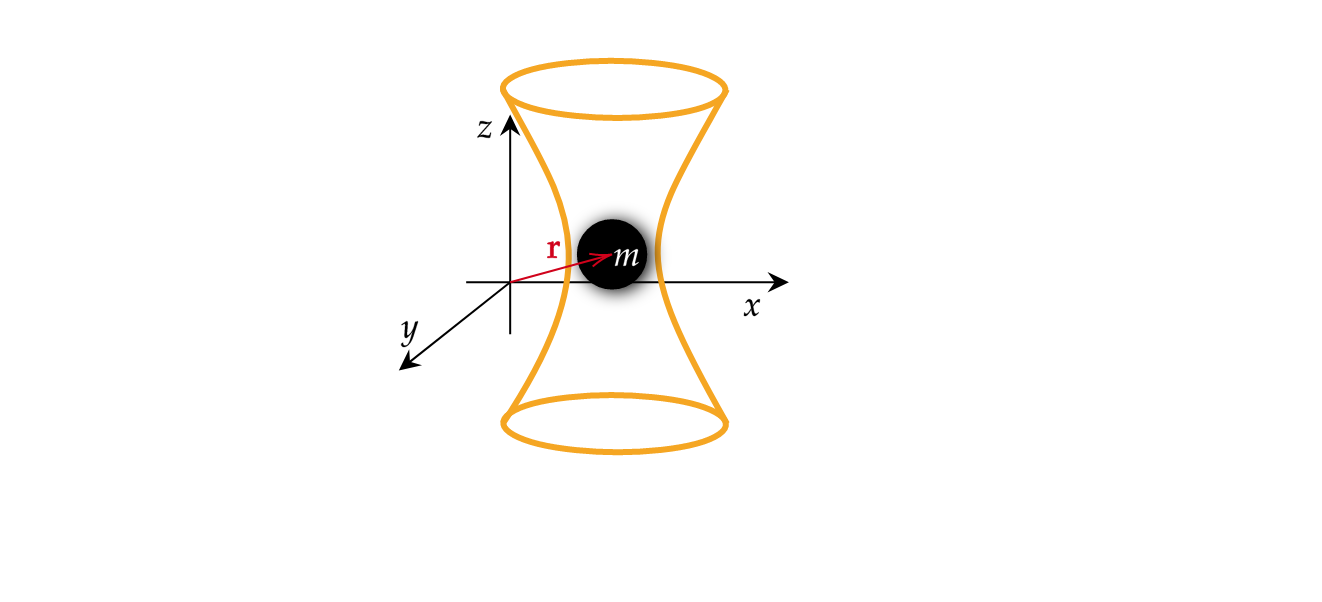}    
    \caption{Particle of mass $m$ in a three dimensional optical trap.}
   \label{fig:optrap}
\end{figure}

We now consider fully decoupled linear stochastic systems (i.e. where $\mathbf{A}$ is a diagonal matrix). A practical example of a three-dimensional linear decoupled process corresponds to the simplified version of the mathematical description of an optical trap shown in Figure \ref{fig:optrap}. The model consists of a set of three independent overdamped Langevin equations \cite{pesce2020optical} given by equation \eqref{application2}. Here, $x$ and $y$ represent the position of the particle in the plane perpendicular to the beam propagation direction and $z$ represents the position of the particle along the propagation direction. The stiffnesses of the trap in each of these directions are $\kappa_x$, $\kappa_y$ and $\kappa_z$, respectively. $\gamma$ is the particle friction coefficient. $\xi_1,\xi_2$ and $\xi_3$ are independent delta-correlated noises, i.e. $\langle \xi_i(t)\rangle=0$, $\langle\xi_i(t)\xi_i(t^\prime)\rangle=2D_{ii}\delta(t-t^\prime)$ and $\langle\xi_i(t)\xi_j(t^\prime)\rangle=0 \quad\forall i\neq j$ with $i=1,2,3$.
\begin{equation}
\begin{bmatrix}
\dot{x}(t) \\ \dot{y}(t) \\ \dot{z}(t)
\end{bmatrix}=\begin{bmatrix}
-\frac{\kappa_x}{\gamma} & 0 & 0 \\
0 & -\frac{\kappa_y}{\gamma} & 0 \\
0 & 0 & -\frac{\kappa_z}{\gamma}
\end{bmatrix}\begin{bmatrix}
x(t) \\ y(t) \\ z(t)
\end{bmatrix}+\begin{bmatrix}
\xi_1(t) \\ \xi_2(t) \\ \xi_3(t)
\end{bmatrix}. \label{application2}
\end{equation}

\begin{figure}[h]
    \centering
    \includegraphics[trim={1cm 0cm 1cm 0cm},clip,width=\columnwidth]{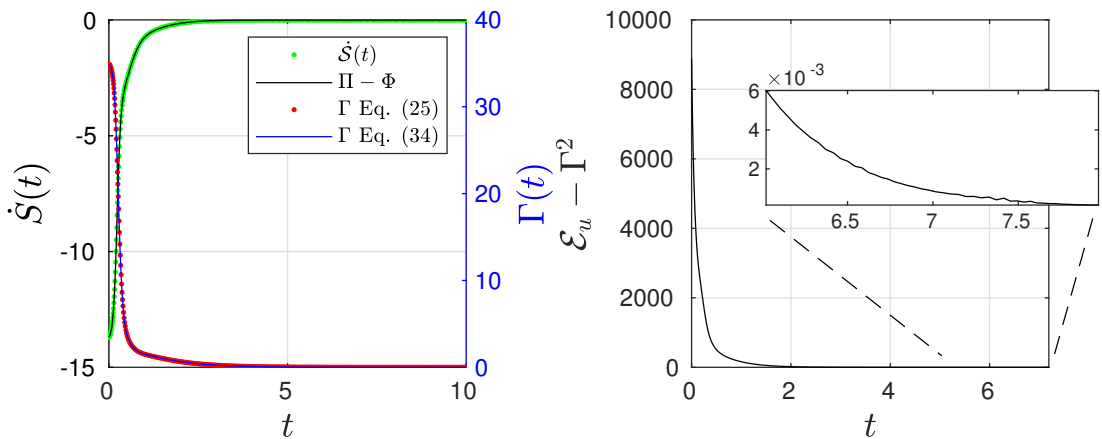}
    \caption{Computational experiment of a three-dimensional optical trap using $\kappa_{x}=10,\kappa_{y}=3$, $\kappa_{z}=1,\gamma=1$, $x(0)=1$, $y(0)=0.1$, $z(0)=0.5$, $\boldsymbol{\Sigma}=0.1\mathbf{I}$ and $\mathbf{D}=0.01\mathbf{I}$ where $\mathbf{I}$ is the identity matrix.}
    \label{fig:ER3exp1}
\end{figure}
Since \eqref{application2} is a fully decoupled linear stochastic model, it permits us to show the applicability of Relation \ref{th3}. To this end, in the left plot of Figure \ref{fig:ER3exp1}, we show $\dot{S}$, $\Pi-\Phi$, and $\Gamma$ computed from equations \eqref{Erate}, \eqref{SPeq} minus \eqref{SFeq}, and \eqref{defirate}-\eqref{eqcol1}, respectively. In Figure \ref{fig:ER3exp1} we can see that both entropy rate $\dot{S}\to 0$ and information rate $\Gamma\to 0$ as the system goes to equilibrium. The right plot of Figure \ref{fig:ER3exp1}, a plot depicting the value of the information rate upper bound $\mathcal{E}_u$ (Relation \ref{mainth}) minus $\Gamma^2$ showing $\mathcal{E}_u\to0$ at equilibrium. The exact value of $\mathcal{E}_u(t)$ in equilibrium is
\begin{equation}
    \lim_{t\to\infty}\mathcal{E}_u=\Tr(\boldsymbol{\Sigma}^{-1}(\infty))\Pi(\infty)\Tr(\mathbf{D})-2g(\mathbf{H}(\infty))=0, \label{eqEulimita}
\end{equation}
where 
\begin{equation}
    \boldsymbol{\Sigma}^{-1}(\infty)=-\mathbf{D}^{-1}\mathbf{A}, \quad\Pi(\infty)=0.\quad \mathbf{H}(\infty)=0. \label{pitf}
\end{equation}
Equation \eqref{pitf} applies only to systems with diagonal and stable $\mathbf{A}$ (proof at \ref{sec:Equlimit}). Since $\Pi\to0$, any decoupled linear system is reversible at equilibrium. 
\subsection{Higher order systems and the upper bounds of information rate $\Gamma$}\label{sec:hos}
As discussed in Section \ref{sec:ERandIL}, it is possible to relate $\dot{S}$, $\Pi$ and $\Gamma$ using the inequalities from Relations \ref{mainth} and \ref{mainth2}. These inequalities become highly relevant when the order of the stochastic models increase, for instance, when using toy models in control engineering scenarios \cite{bechhoefer2021control}. In this section, we take the case when $\mathbf{A}$ is a randomly chosen Hurwitz matrix whose size varies from $2-50$, i.e. we choose linear stochastic systems that contain from $2$ to $50$ random variables. 
\begin{figure}[h]
    \centering
    \begin{subfigure}[b]{0.6\textwidth}
    \centering
    \includegraphics[trim={0.7cm 0.5cm 0cm 1cm},clip,width=\columnwidth]{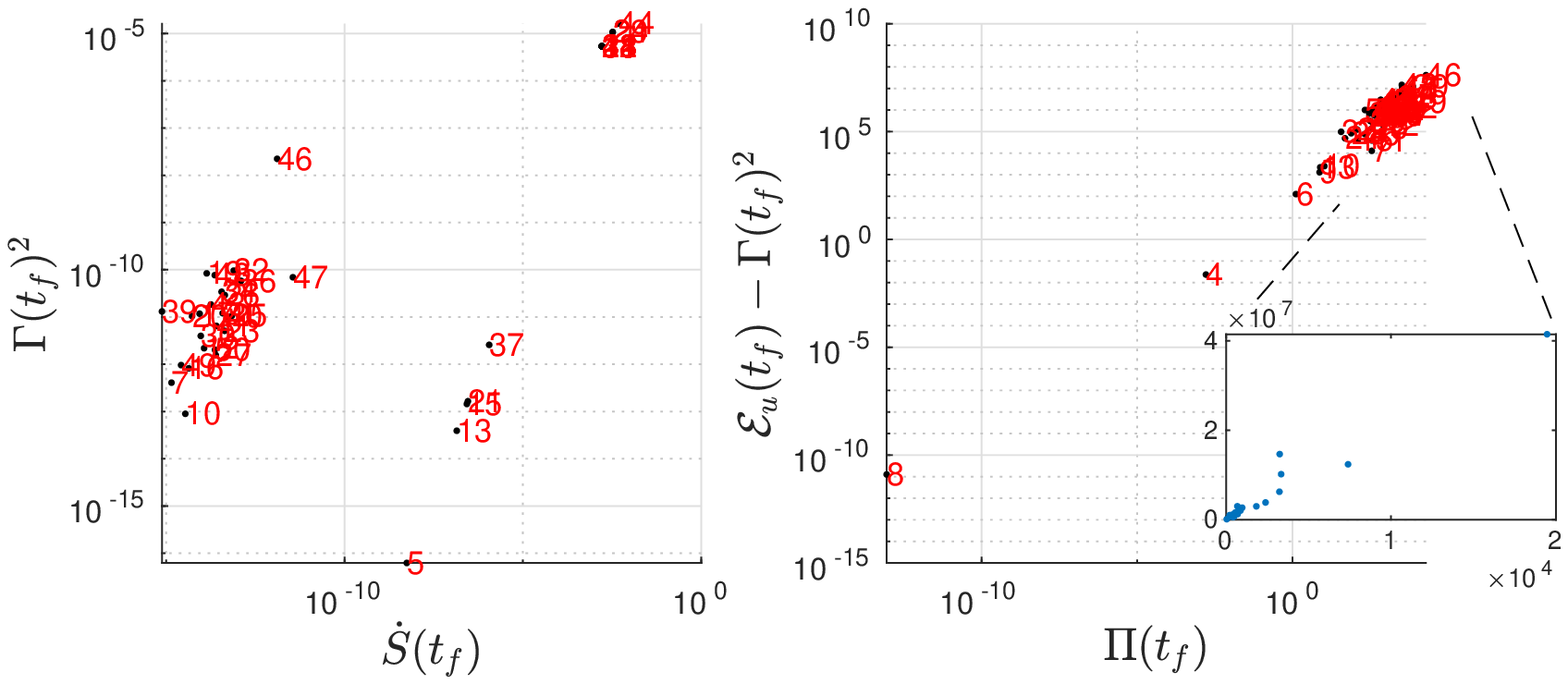}\vspace{-0.5cm}
    \caption{$\mathbf{A}$ is non diagonal.}\label{fig:multidimensionala}   
    \end{subfigure}    
    \begin{subfigure}[b]{0.6\textwidth}
     \centering
    \includegraphics[trim={0.7cm 0.5cm 1cm 1cm},clip,width=\columnwidth]{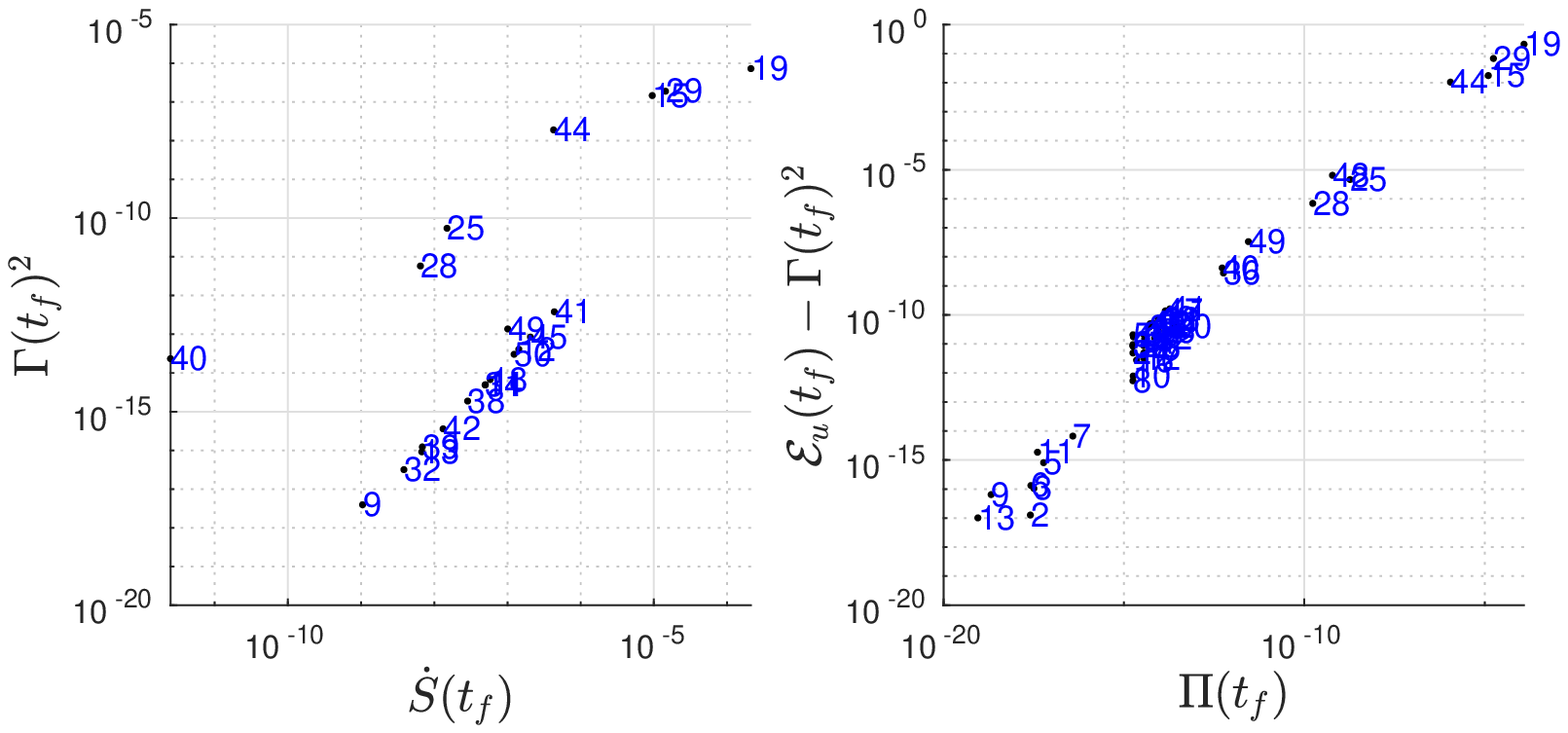} \vspace{-0.8cm}
    \caption{$\mathbf{A}$ is diagonal.}\label{fig:multidimensionalb}   
    \end{subfigure}
\caption{Entropy rate $\dot{S}(t_f)$ vs the square of information rate $\Gamma(t_f)^2$ and the values of $\mathcal{E}_u(t_f)-\Gamma(t_f)^2$ vs $\Pi(t_f)$. 
The simulations use randomly chosen stable linear systems from order $n=2$ to $n=50$. The red numbers indicate the order of the system and its position the value at $t_f=300$. In all simulations the system starts out of the equilibirum with initial conditions $\boldsymbol{\mu}=[1,1,\dots,1]^\top,\boldsymbol{\Sigma}=0.1\mathbf{I}$ and $\mathbf{D}=0.01\mathbf{I}$ where $\mathbf{I}$ is the identity matrix varying from size 2 to 50.}\label{fig:multidimensional}   
\end{figure}
Figure \ref{fig:multidimensional} shows the phase portrait of $\Gamma(t_f)^2$ vs $\dot{S}(t_f)$, and the phase portrait of $\mathcal{E}_u(t_f)-\Gamma(t_f)^2$ vs $\Pi(t_f)$. Figure \ref{fig:multidimensional} is also separated in sub-figures \ref{fig:multidimensionala} and \ref{fig:multidimensionalb} showing the cases when matrix $\mathbf{A}$ is diagonal and non diagonal, respectively. Note, $t_f$ refers to the time close enough the system's equilibrium; in our simulations $t_f=300$. The phase portraits contain numbers to indicate the value at $t=t_f$; the number also indicates the order of the stochastic system.

Regarding the portraits of $\Gamma(t_f)^2$ vs $\dot{S}(t_f)$, for every Hurwitz $\mathbf{A}$ (i.e. diagonal and non diagonal matrix with negative real part eigenvalues) $\lim_{t\to\infty}\mathcal{E}(t)=\lim_{t\to\infty}\dot{S}(t)=0$ as expected. Meanwhile, for the same processes $\Pi> 0$ at equilibrium (see equation \eqref{limitpi}). When looking at the phase portraits of $\mathcal{E}_u(t_f)-\Gamma(t_f)^2$ vs $\Pi(t_f)$, we see that  $\mathcal{E}_u(t_f)>0$ when $\mathbf{A}$ is non diagonal due to $\Pi(t_f)>0$ for some $t$ (see Figure \ref{fig:multidimensionala}). On the other hand, as demonstrated in Equations \eqref{eqEulimita} and \eqref{pitf} $\mathcal{E}_u(t_f)=\Pi(t_f)=0$. Again, meaning that every fully decoupled linear system is reversible at equilibrium. 

\subsection{Abrupt events analysis}\label{sec:aea}
When the temperature changes abruptly in a system like \eqref{bmotiona}, the value of $D$ (noise amplitude) is affected, contributing to the uncertainty in the control of the particle's position. To bring light to the analysis and study of abrupt events, we use our toy models and simulate an abrupt change in the system's temperature by using the following impulse like function for the $ii$-element of the noise amplitude matrix $\mathbf{D}$ and on the input function $u(t)$
\begin{eqnarray}
    D_{ii}(t)=D_0+\frac{1}{|a|\sqrt{\pi}}e^{-\left(\frac{t-t_p}{a}\right)^2},
    \label{eqD}\\
     u(t)=\frac{1}{|a|\sqrt{\pi}}e^{-\left(\frac{t-t_p}{a}\right)^2}.   \label{equ}
\end{eqnarray}
Here, the second term on RHS of \eqref{eqD} and \eqref{equ} takes a non-zero value for a short time interval around $t_p$ and $a$ changes the amplitude of the impulses. 
\subsubsection{Harmonically bound particle}
\begin{figure}[h!]
    \centering
    \begin{subfigure}[b]{0.7\columnwidth}
    \centering
    \includegraphics[trim={1cm 0cm 1.2cm 0cm},clip,width=\columnwidth]{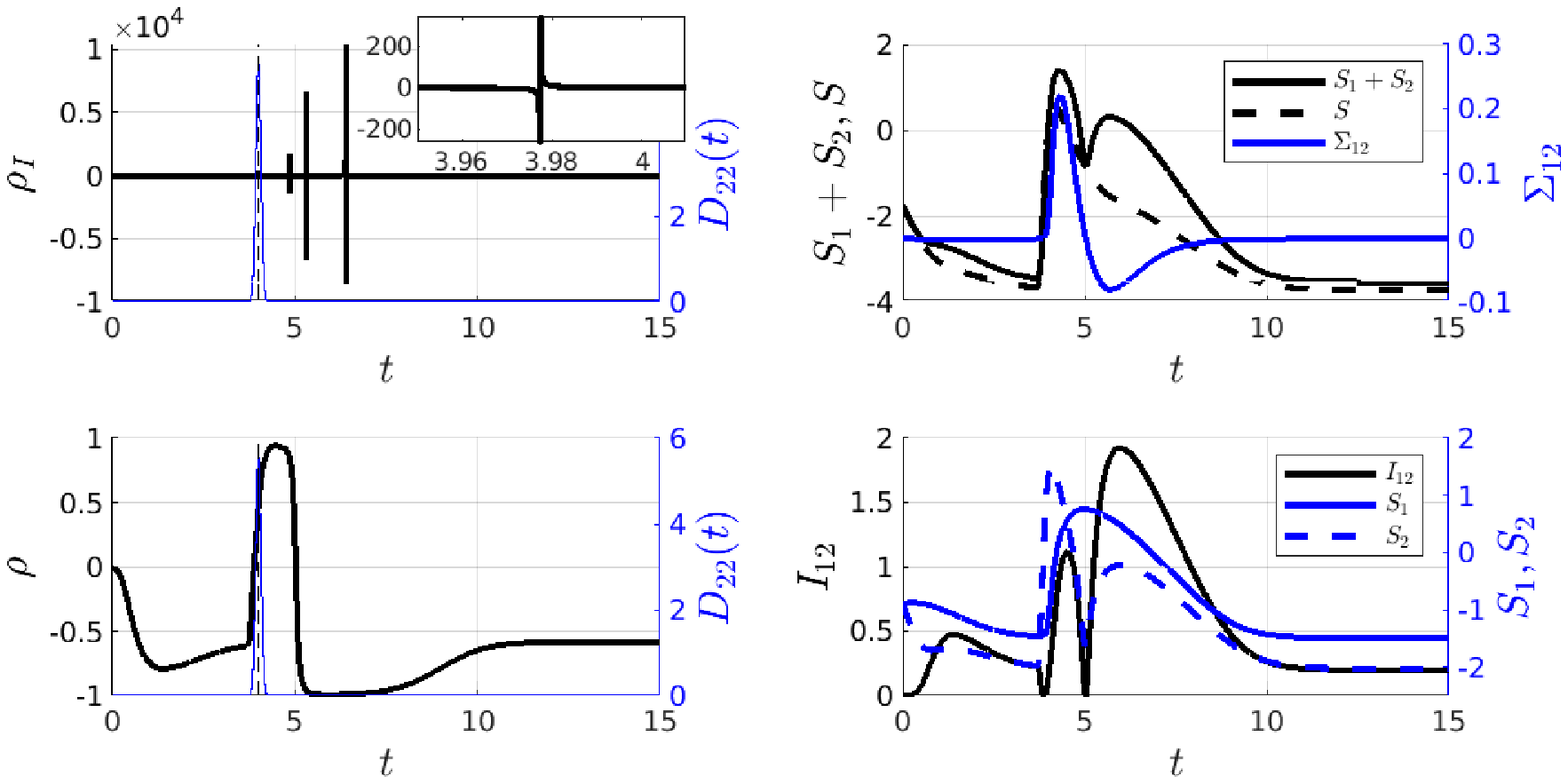}\vspace{-0.7cm}
    \caption{}\label{fig:AEexp1a}
    \end{subfigure}
    \begin{subfigure}[b]{0.7\columnwidth}
    \centering
    \includegraphics[trim={1cm 0cm 1.5cm 0cm},clip,width=\columnwidth]{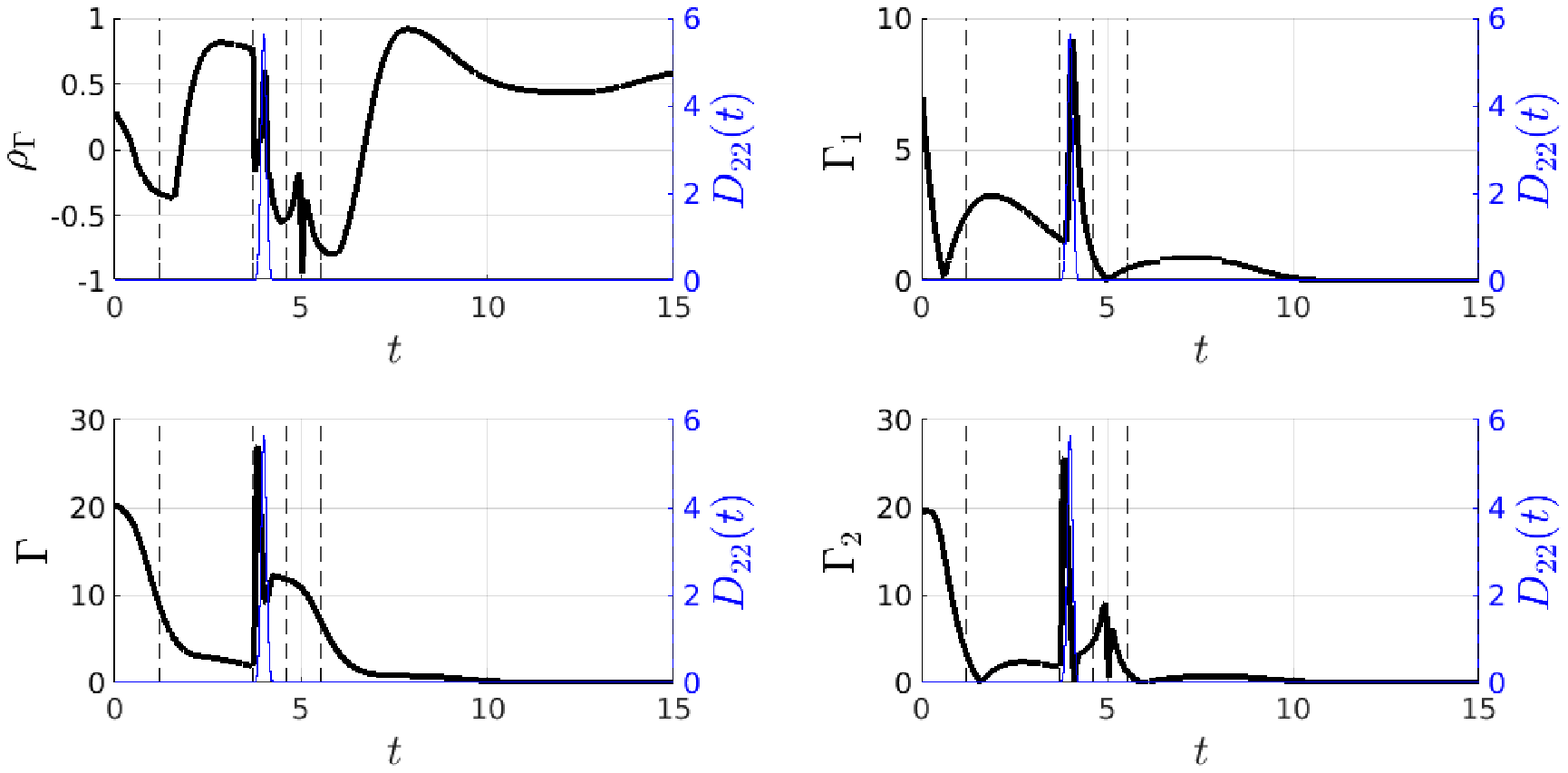}\vspace{-0.7cm}
    \caption{}\label{fig:AEexp1b}
    \end{subfigure}
    \begin{subfigure}[b]{0.7\columnwidth}
    \centering
    \includegraphics[trim={0.5cm 0cm 1.5cm 0cm},clip,width=\columnwidth]{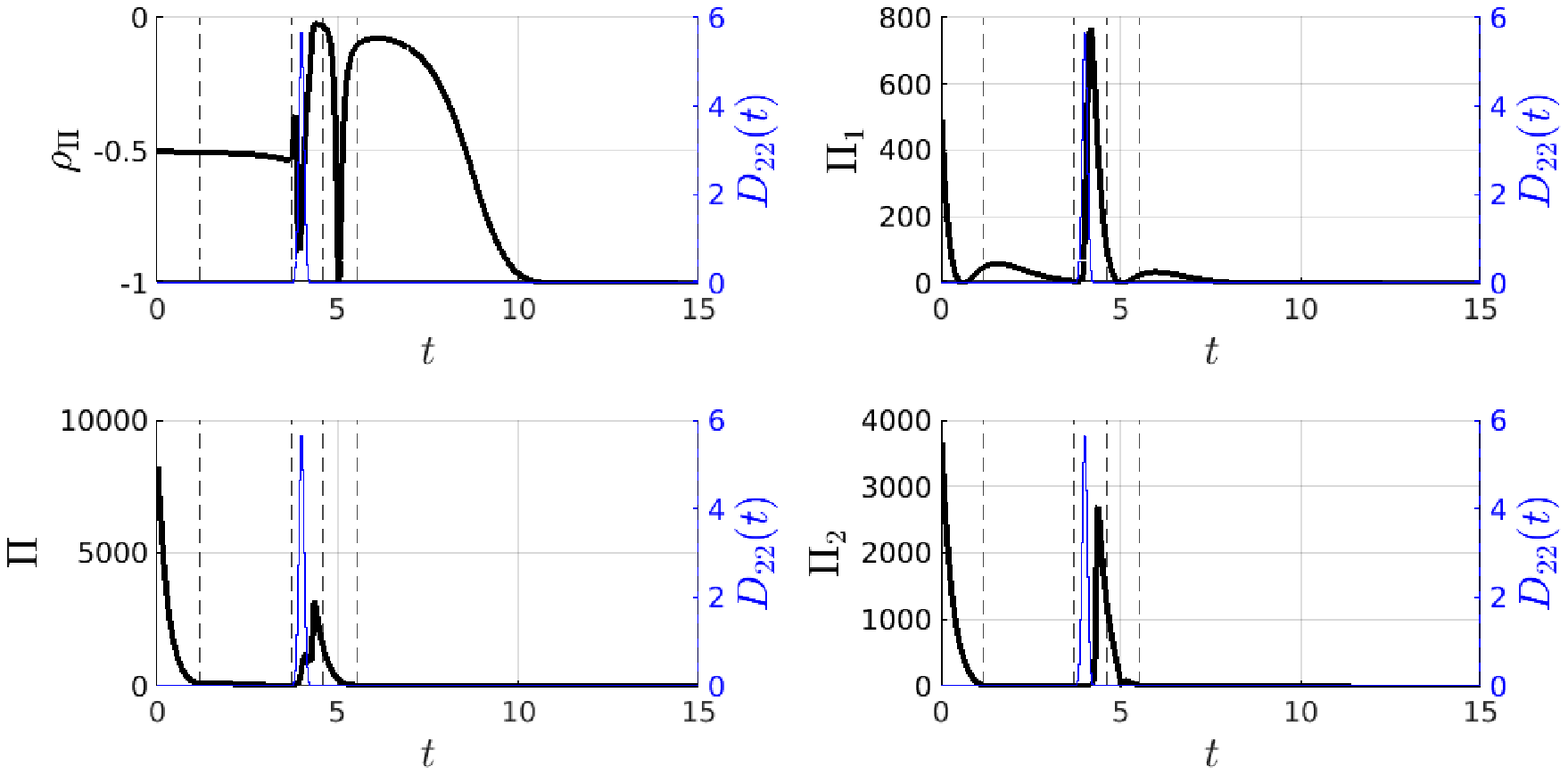}\vspace{-0.7cm}
    \caption{}\label{fig:AEexp1c}
    \end{subfigure}    
    \caption{Abrupt event numerical experiment $u(t)=0,D_{11}(t)=0.001$ and $D_{22}(t)=0.001+\frac{1}{|0.1|\sqrt{\pi}}e^{-\left(\frac{t-4}{0.1}\right)^2}$. The rest of the simulation parameters are $\omega=1,\gamma=2$, $\mu_1(0)=0.5,\mu_2(0)=0.7$, and $\boldsymbol{\Sigma}(0)=0.01\mathbf{I}$ where $\mathbf{I}$ is the identity matrix.}
    \label{fig:AEexp1}
\end{figure}

\begin{figure}[h!]
    \centering
    \begin{subfigure}[b]{0.7\columnwidth}
    \centering
    \includegraphics[trim={1cm 0cm 1.5cm 0cm},clip,width=\columnwidth]{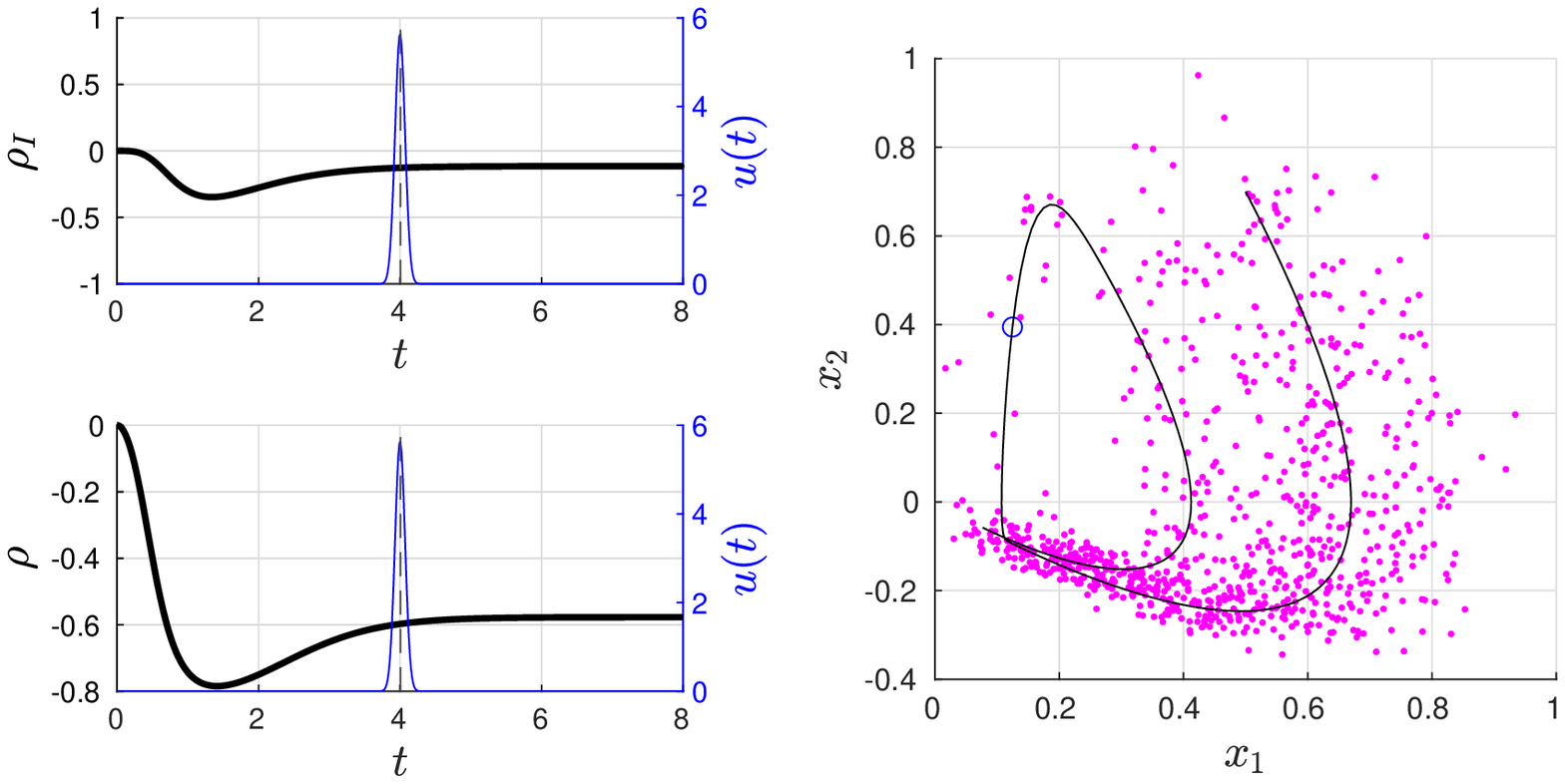}\vspace{-0.7cm}
    \caption{}\label{fig:AEexp2a}
    \end{subfigure}
    \begin{subfigure}[b]{0.7\columnwidth}
    \centering
    \includegraphics[trim={1cm 0cm 1.5cm 0cm},clip,width=\columnwidth]{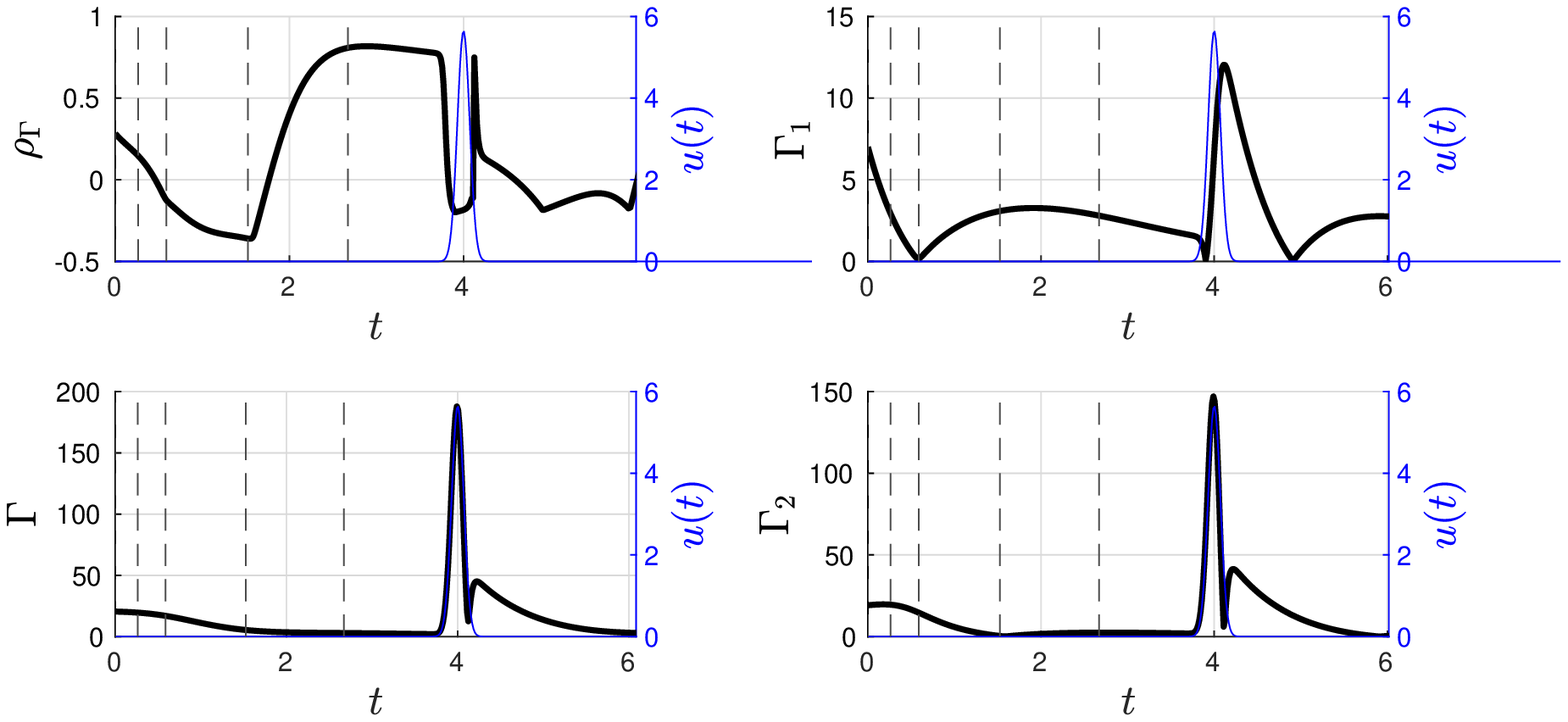}\vspace{-0.7cm}
    \caption{}\label{fig:AEexp2b}
    \end{subfigure}
    \begin{subfigure}[b]{0.7\columnwidth}
    \centering
    \includegraphics[trim={1cm 0cm 1.5cm 0cm},clip,width=\columnwidth]{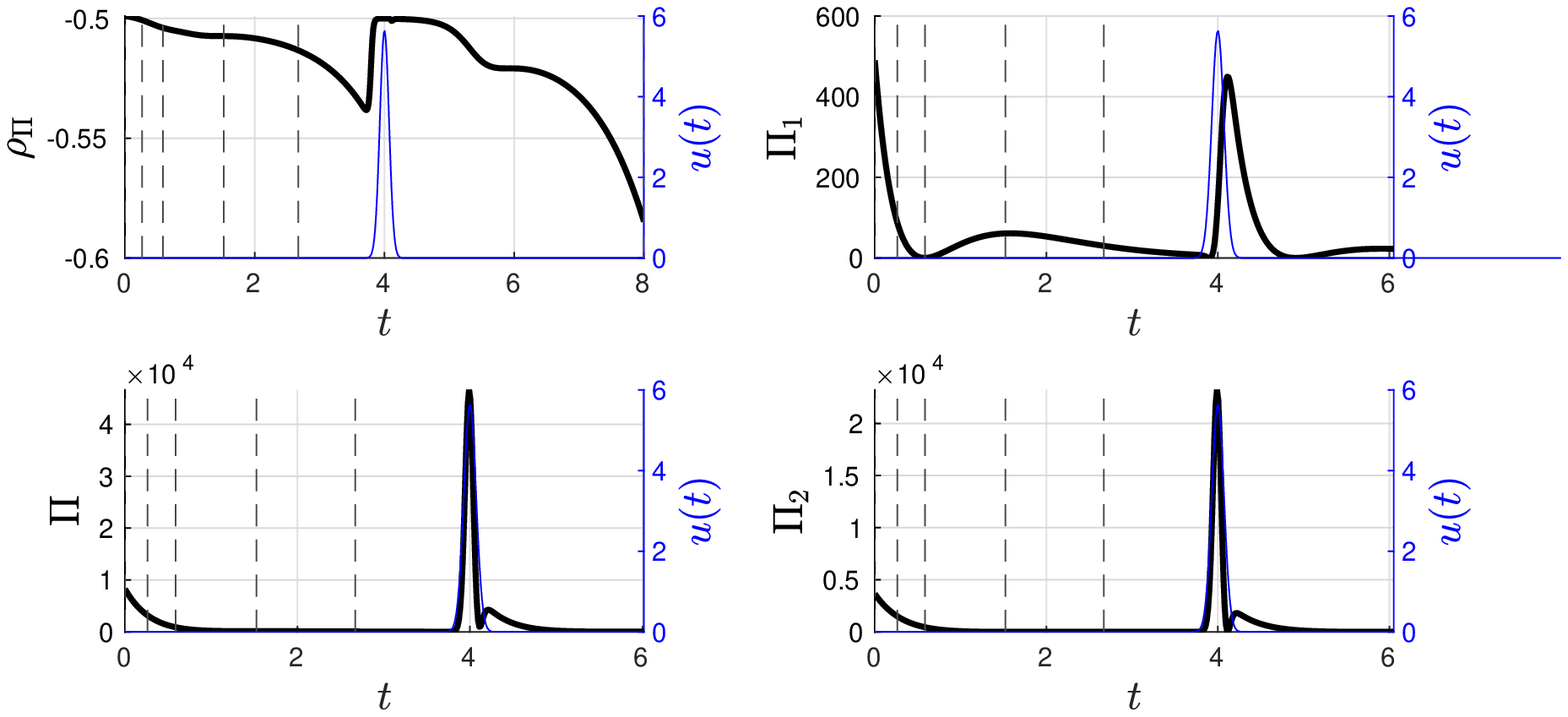}\vspace{-0.7cm}
    \caption{}\label{fig:AEexp2c}
    \end{subfigure}    
    \caption{Abrupt event numerical experiment $D_{22}(t)=D_{11}(t)=0.001$ and $u(t)=\frac{1}{|0.1|\sqrt{\pi}}e^{-\left(\frac{t-4}{0.1}\right)^2}$. The rest of the simulation parameters are $\omega=1,\gamma=2$, $\mu_1(0)=0.5,\mu_2(0)=0.7$, and $\boldsymbol{\Sigma}(0)=0.01\mathbf{I}$ where $\mathbf{I}$ is the identity matrix.}
    \label{fig:AEexp2}
\end{figure}

Again, we start our analysis by considering system \eqref{bmotiona}. Figures \ref{fig:AEexp1} and \ref{fig:AEexp2} show the computer simulation results and time evolution of abrupt events in $\mathbf{D}$ and $u$, respectively. The noise amplitude is perturbed via the element $D_{22}$ of the matrix $\mathbf{D}$ and the input force $u$ only affects the state $x_2$. Figures \ref{fig:AEexp1} and \ref{fig:AEexp2} are divided in three panels, \ref{fig:AEexp1a}/\ref{fig:AEexp2a} which includes the time evolution of the correlation coefficients $\rho$ and $\rho_I$ (Figure \ref{fig:AEexp2a} also includes the phase portrait of $x_1$ vs $x_2$); \ref{fig:AEexp1b}/\ref{fig:AEexp2b} shows the time evolution of $\rho_\Gamma,\Gamma,\Gamma_1$ and $\Gamma_2$; \ref{fig:AEexp1c}/\ref{fig:AEexp2c} the time evolution of $\rho_\Pi,\Pi,\Pi_1$ and $\Pi_2$. 
From Figure \ref{fig:AEexp1}, the coefficient $\rho_I$ is the most sensitive to noise amplitude perturbations, as it shows an abrupt change before the peak of the perturbation at $t=4$ followed by a couple of extra abrupt changes at $t \geq 5$. To explain the sensitiveness of $\rho_I$, in \ref{fig:AEexp1a}, we have included time evolution of $S_1+S_2,S,\Sigma_{22},I_{12},S_1$ and $S_2$ instead of a phase portrait as in \ref{fig:AEexp2a}. The divergences in $\rho_I$ can be explained if we notice that the values of entropy (specially $S_2$) go from negative to positive values after the perturbation occurs. Let us recall that differential entropy $S$ can be negative and it is understood as a relative privation of information. When negative, its value means we have less disorder/uncertainty  or more information/order. The increment in temperature given by the perturbation increases the level of uncertainty in the system and thus entropy, then when it decreases entropy decreases as well. Such changes lead to $S,S_1+S_2$ and $I_{12}$ to be zero provoking divergences in $\rho_I$ at different instants of time. Meanwhile, since it precedes the aforementioned perturbation (see Figure \ref{fig:AEexp1b}), the value of $\Gamma$ predicts the abrupt event (corroborating the previous results shown in \cite{guel2021information}). Regarding the perturbation in $u(t)$ shown in Figure \ref{fig:AEexp2}, the coefficients $\rho$ and $\rho_I$ are no longer useful because they are not sensitive to changes in the mean value of the PDF (see Figure \ref{fig:AEexp2a}). In contrast, an abrupt event in the mean value is well captured by $\rho_\Gamma$ and $\rho_\Pi$. 

Figures \ref{fig:AEexp2b} and \ref{fig:AEexp2c} show that the values of $\rho_\Gamma$ and $\rho_\Pi$ change abruptly at the time $t=4$ when perturbation occurs. Figure \ref{fig:AEexp2b} presents negative $\rho_\Gamma$ at $t\approx 4$ due to the large difference between $\Gamma_2$ and $\Gamma_1$. For the similar reasons, $\rho_\Pi$ also presents a high decrement at $t\approx 4$. Here, the coefficients are able to detect the perturbation over the mean value but they are no longer able to predicted it.

\subsubsection{Controllable canonical form}
\begin{figure}[h]
    \centering
    \begin{subfigure}[b]{0.5\columnwidth}
    \centering
    \includegraphics[trim={1cm 0cm 0.5cm 0cm},clip,width=\columnwidth]{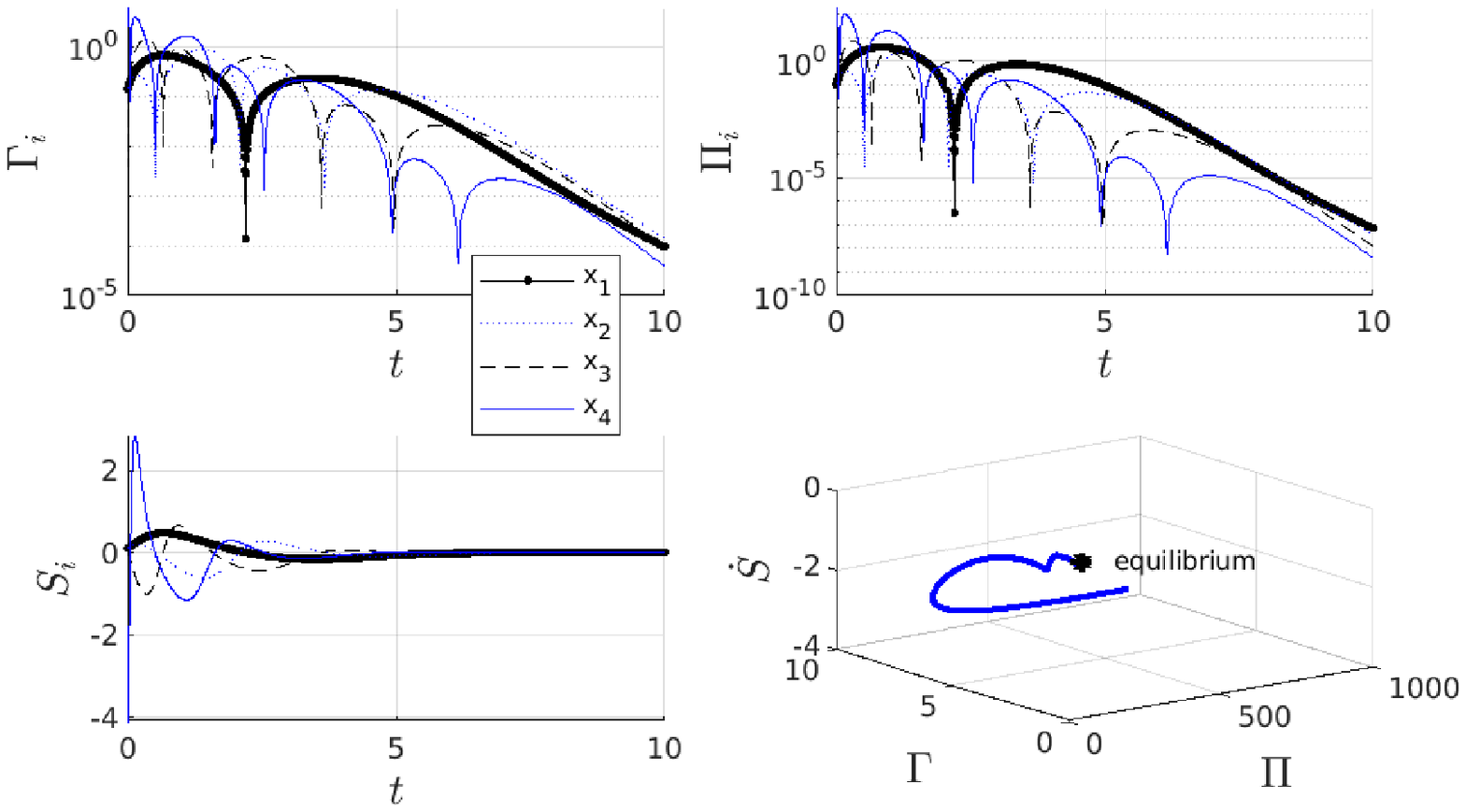}  
    \caption{}    \label{fig:cforma}
    \end{subfigure}
    \begin{subfigure}[b]{0.49\columnwidth}
    \centering
    \includegraphics[trim={1cm 1cm 0.8cm 0.8cm},clip,width=0.6\columnwidth]{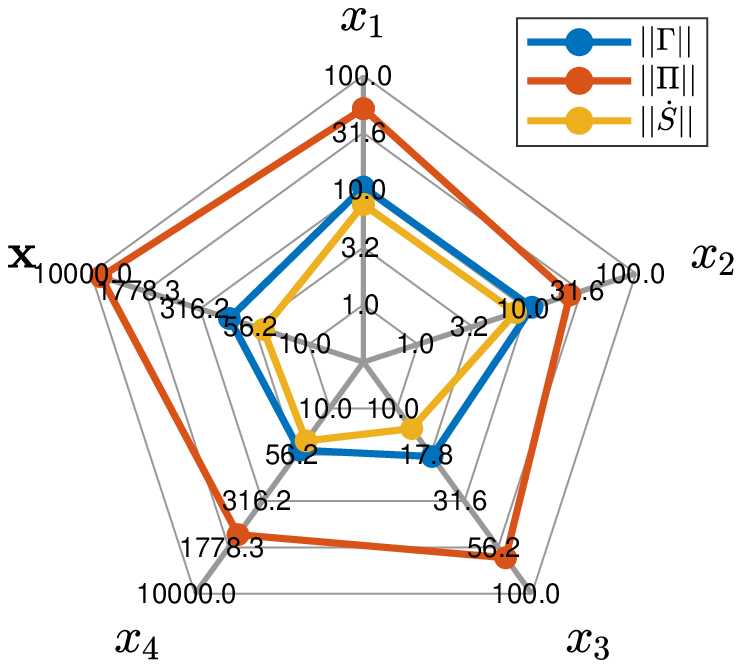}
    \caption{}    \label{fig:cformb}
    \end{subfigure}
    \caption{Behaviour with no abrupt event $D_{ii}(t)=D_0=0.01,u(t)=0$. The rest of the simulation parameters are $\mu(0)=[0,0,0,1]^\top$ and $\boldsymbol{\Sigma}(0)=0.1\mathbf{I}$ where $\mathbf{I}$ is the identity matrix.}
    \label{fig:cform}
\end{figure}

\begin{figure}[h]
    \centering
    \begin{subfigure}[b]{0.5\columnwidth}
    \centering
    \includegraphics[trim={1cm 0cm 1cm 0cm},clip,width=\columnwidth]{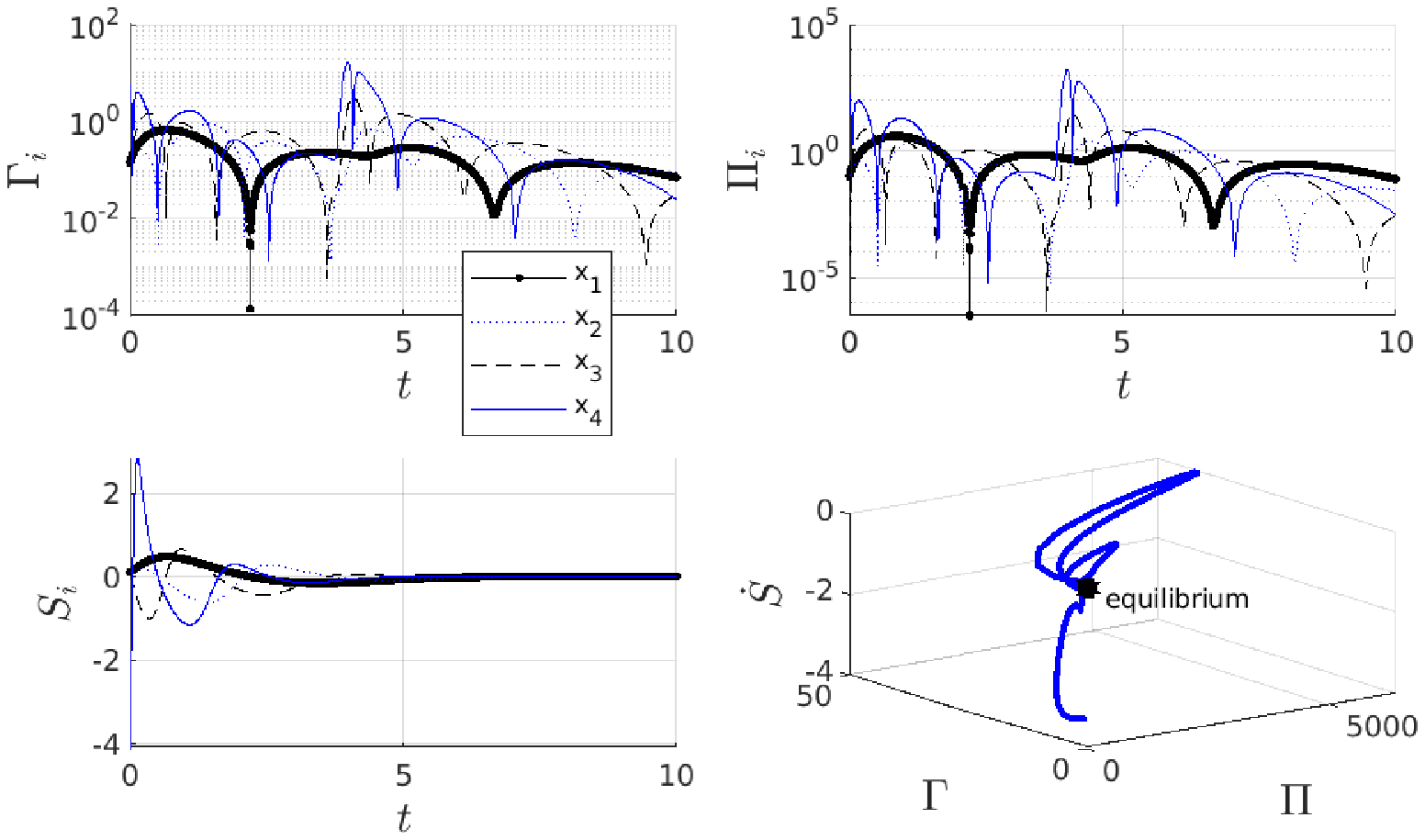}  
    \caption{}    \label{fig:cform3a}
    \end{subfigure}
    \begin{subfigure}[b]{0.49\columnwidth}
    \centering
    \includegraphics[trim={1cm 1cm 0.8cm 1cm},clip,width=0.6\columnwidth]{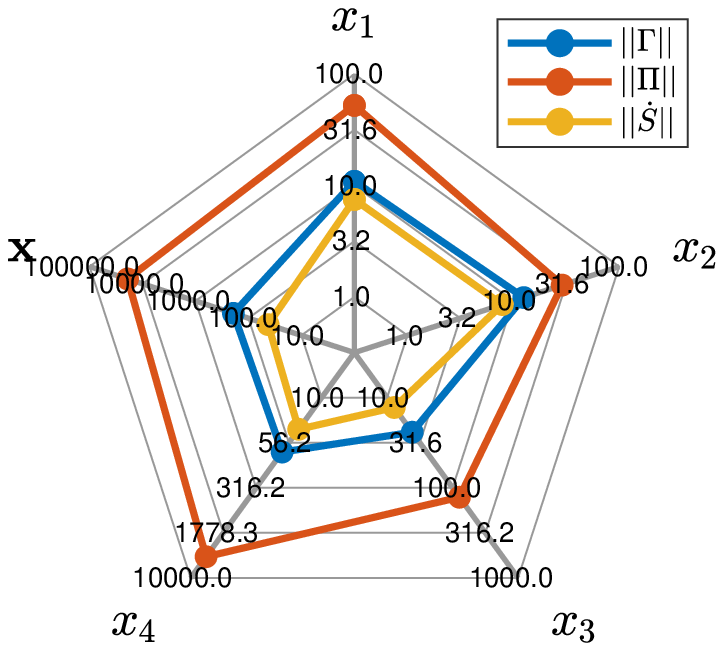}
    \caption{}    \label{fig:cform3b}
    \end{subfigure}
    \caption{Behaviour with abrupt event $D_{ii}(t)=D_0=0.01,u(t)=\frac{1}{|0.1|\sqrt{\pi}}e^{-\left(\frac{t-t_p}{0.1}\right)^2}$ with $t_p=4$. The rest of the simulation parameters are $\mu(0)=[0,0,0,1]^\top$ and $\boldsymbol{\Sigma}(0)=0.1\mathbf{I}$ where $\mathbf{I}$ is the identity matrix.}
    \label{fig:cform3}
\end{figure}

\begin{figure}[h]
    \centering
    \includegraphics[trim={1.5cm 0.5cm 1cm 0cm},clip,width=\columnwidth]{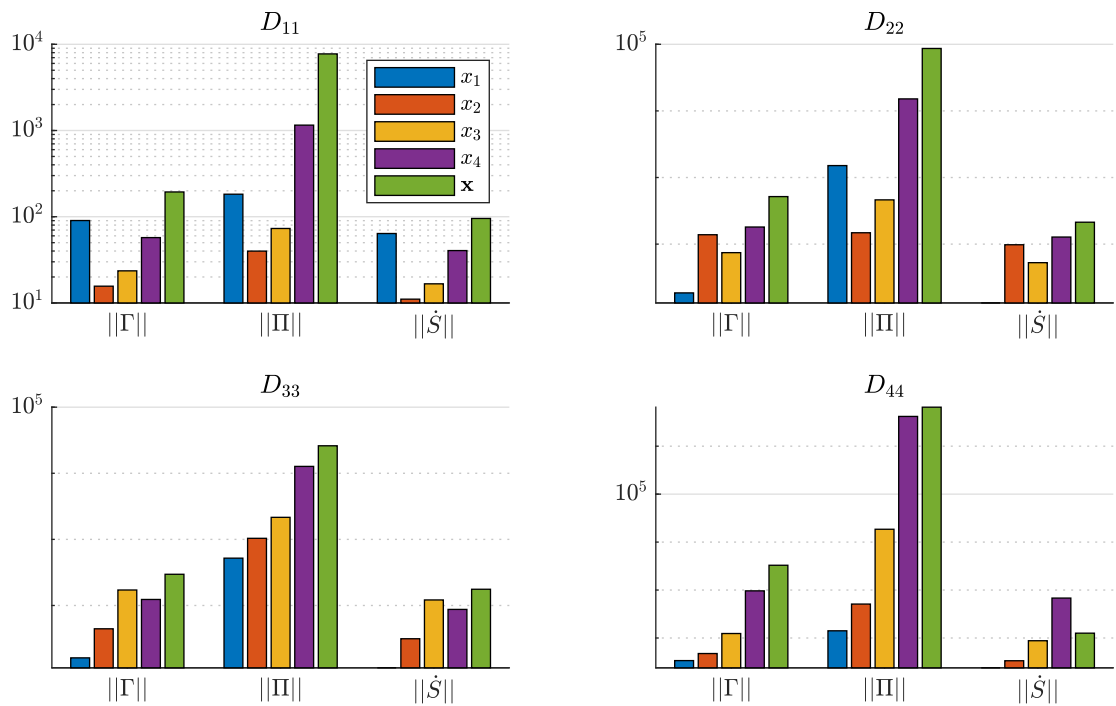} 
    \caption{Abrupt event analysis using the norms of $\Pi$, $\dot{S}$, $\Gamma$ at the marginal PDF $p(x_i,t)$ and the joint PDF $p(\mathbf{x};t)$. Each plot depicts a perturbation on a given $D_{ii}$ according to \eqref{eqD} at $t_p=6$ with $a=0.1$ and $D_0=0.01$ in system \eqref{eqcform}. The rest of the simulation parameters are $\mu(0)=[0,0,0,1]^\top$ and $\boldsymbol{\Sigma}(0)=0.1\mathbf{I}$ where $\mathbf{I}$ is the identity matrix.}
    \label{fig:cform2}
\end{figure}

The analysis of abrupt events in high order systems can be done by an offline method, such as the application of the Euclidean norm to each marginal or joint information rate/entropy production of the random variables in the system. Recall that the Euclidean norm of any time dependant function $\vartheta(t)$ is defined as follows
\begin{equation}
    \|\vartheta(t)\|:=\left(\int_0^{t_f}\vartheta(\tau)^2d\tau\right)^{\frac{1}{2}}. \label{normdef}
\end{equation}
As a demonstration, using Equation \eqref{normdef}, here we study the effects of abrupt events in the noise amplitude matrix $\mathbf{D}(t)$ and the force input $u(t)$ of the popular controllable canonical form of the state-space realization of a linear system given by 
\begin{equation}
\dot{\mathbf{x}}(t)=
\begin{bmatrix}
0 & 1 & 0 & \cdots & 0 \\
0 & 0 & 1 & \cdots & 0 \\
\vdots & \vdots & \vdots & \ddots & \vdots \\
0 & 0 & 0 & \cdots & 1 \\
-d_n & -d_{n-1} & -d_{n-2} & \cdots & -d_1
\end{bmatrix}\mathbf{x}(t)
+\begin{bmatrix}
0\\ \vdots \\ 1
\end{bmatrix}u(t)+\boldsymbol{\xi}(t). \label{eqcform}
\end{equation}
Here, $\mathbf{x}:=[x_1,x_2,\dots,x_n]^\top\in\mathbb{R}^n$, $u\in\mathbb{R}$ and $\boldsymbol{\xi}:=[\xi_1,\xi_2,\dots,\xi_n]^\top\in\mathbb{R}^n$ is a vector of random variables with $\langle \xi_i(t)\rangle=0$, and $\langle\xi_i(t)\xi_j(t^\prime)\rangle=2D_{ij}\delta(t-t^\prime)$. Model \eqref{eqcform} provides us with a structure where the input enters a chain of integrators making it to move every state in the Langevin equation (i.e. they are fully controllable). We consider the case when \eqref{eqcform} is of $4$th order. The values of the parameters are $\mathbf{d}_4=[d_1,\dots,d_n]^\top=[-1.5165,-5.2614,-6.7985, -4.2206]^\top.$ In Figures \ref{fig:cform} to \ref{fig:cform2}, we use the notation $x_i$ $\forall i=1,2,3,4$ and $\mathbf{x}$ to refer to the values of $\Pi,\dot{S}$ and $\Gamma$ computed from marginal PDF $p(x_i,t)$ and from the joint PDF $p(\mathbf{x};t)$, respectively.

Figure \ref{fig:cforma} depicts the time evolution of $\Gamma_i,\Pi_i$ and $\dot{S}_i$ $\forall i=1,2,3,4$. It also includes the time evolution of the three dimensional space $(\Gamma,\Pi,\dot{S})$. Figure \ref{fig:cformb} shows the norms of $\Gamma_i,\Pi_i,\dot{S}_i$ $\forall i=1,2,3,4$ and $\Gamma,\Pi,\dot{S}$ in the form of a spider plot. For instance, the value of the norm of the information rate $\Gamma$ computed from the joint PDF $p(\mathbf{x};t)$ and from the marginal PDF $p(x_2,t)$ is $||\Gamma||\approx 10000$ and $||\Gamma_2||\approx 31.6$, respectively. As we can see, the effects on $\Gamma_i,\Pi_i$ and $\dot{S}_i$ $\forall i=1,2,3,4$ by the random variables is hierarchical with regards to their amplitude (for example $|\dot{S}_4|>|\dot{S}_3|>|\dot{S}_2|>|\dot{S}_1|$ at almost all the time) and the equilibrium of $(\Gamma,\Pi,\dot{S})$ is $(0,0,0)$. 
When we add a perturbation in $u$ which affects directly to $x_4$ (see Equation \eqref{eqcform}), we obtain the results shown in Figure \ref{fig:cform3}. Such an abrupt event causes an notable increment in the norms of the states (See Figure \ref{fig:cform3b}) which still maintains the same hierarchical order in the states ($x_4$ is the most affected in comparison with $x_1$) due to the system's structure as expected. The direct effect of the abrupt event on each variable's time evolution is shown in Figure \ref{fig:cform3a}. Recall that $u$ is applied directly to $x_4$. Again, $\dot{S}$ is unperturbed since the event affects only the mean value of the PDF.
On the other hand, if we separately include a perturbation in each element $D_{ii}$ of the noise matrix $\mathbf{D}$, similar results occur. Figure \ref{fig:cform2} illustrates the norms of $\Gamma,\Pi,\dot{S}$ in the form of bar plots (after applying perturbations to each $D_{ii}$). The plots indicate a domino effect in the marginal PDFs as follow. When only $D_{11}$ is perturbed no clear effects can be seen in the rest of the variables but when $D_{33}$ is perturbed $x_3,x_2$ and $x_1$ increase their values. Same happens after perturbing $D_{44}$, again this is due to the structure interaction of the system we are studying. This implies that norms provide an approximate value of the dependence between the variables of the random process.
\subsection{Minimum variability control design}\label{sec:mep}
The design of ``efficient'' machines remains still a common practical problem in different research areas. Note that here, efficiency can be related to the existence of minimum variability or entropy production as it was highlighted in previous works \cite{kim2017geometric,kim2016geometric,salamon1980minimum,martyushev2006problem}. The generation of efficient processes, as we describe here, can be accomplished by designing optimal protocols through work and heat minimisation \cite{Aurell2011}, the Wasserstein distance \cite{Dechant2019} or inverse engineering (for a complete review, see \cite{Guery2022}) to drive our system while being subject to thermodynamic constraints. In this context, as a final application, here we explore the use of IL to design a classical control technique in a given linear stochastic process. From the field of control engineering, we will take the full-state feedback controller given by 
\begin{equation}
    u(t)=-\mathbf{k}\mathbf{x}(t),
\end{equation}
where $\mathbf{k}\in\mathbb{R}^{1\times n}$. Through this control, we obtain the following closed-loop system
\begin{equation}
\dot{\mathbf{x}}(t)=\mathbf{A}_{cl}\mathbf{x}(t)+\boldsymbol{\xi}(t), \label{clsys}
\end{equation}
where $\mathbf{A}_{cl}=\mathbf{A}-\mathbf{B}\mathbf{k}$. The full-state feedback control permits us to manipulate the system's mean value via changing the eigenvalues of $\mathbf{A}$. As discussed previously, such eigenvalues also modify the time evolution of $\boldsymbol{\Sigma}$. In systems like \eqref{bmotiona}, the value of $\boldsymbol{\Sigma}$ can as well be manipulated by the temperature of the environment whose value is related to the elements $D_{11}$ and $D_{22}$ of the noise amplitude matrix $\mathbf{D}$. 

Taking the aforementioned details into consideration, we propose the following optimisation problems for the design of minimum variability controls
\begin{equation}
\begin{aligned}
\min_{\mathbf{k},\mathbf{D}} \quad & J_1=\int_0^{t_f}\Gamma(\tau)\rmd\tau, \\
\textrm{s.t.} \quad & \dot{\boldsymbol{\mu}}=\mathbf{A}_{cl}\boldsymbol{\mu}\\
  &\dot{\boldsymbol{\Sigma}}= \mathbf{A}_{cl}\boldsymbol{\Sigma}+\boldsymbol{\Sigma}\mathbf{A}_{cl}^\top+2\mathbf{D}   \\
  &\boldsymbol{\mu}(0)=\mathbf{m}, \boldsymbol{\Sigma}(0)=\mathbf{S}\\
  &k_{l,i}\leq k_i \leq k_{u,i},\quad 0\leq D_{ii}\leq D_{\text{max}}\\
  &\forall i=1,2,\dots,n,
\end{aligned}\label{opt1}
\end{equation}
and 
\begin{equation}
\begin{aligned}
\min_{\mathbf{k},\mathbf{D}} \quad & J_2=||\Gamma(t)^2-\Gamma(0)^2||, \\
\textrm{s.t.} \quad & \dot{\boldsymbol{\mu}}=\mathbf{A}_{cl}\boldsymbol{\mu}\\
  &\dot{\boldsymbol{\Sigma}}= \mathbf{A}_{cl}\boldsymbol{\Sigma}+\boldsymbol{\Sigma}\mathbf{A}_{cl}^\top+2\mathbf{D}   \\
  &\boldsymbol{\mu}(0)=\mathbf{m}, \boldsymbol{\Sigma}(0)=\mathbf{S} \\
  &k_{l,i}\leq k_i \leq k_{u,i},\quad 0< D_{ii}\leq D_{\text{max}}\quad \\
  &\forall i=1,2,\dots,n.
\end{aligned}\label{opt2}
\end{equation}
In Equation \eqref{opt1}, $J_1$ is a cost function that considers the minimisation of IL from $t=0$ to $t=t_f$ to obtain the ``minimum'' statistical changes in the given period of time. On the other hand, Equation \eqref{opt2} considers a cost function $J_2$ equal to the norm of $\Gamma(t)^2-\Gamma(0)^2$. The objective of $J_2$ is to keep $\Gamma^2$ constant through time (with the least amount of fluctuations) to approximately follow the ``geodesic'', a problem well described in \cite{kim2016geometric}. Both optimisation problems are subject to the dynamics of the mean and covariance of the PDF given certain initial conditions for them. The problems also consider upper and lower limits to $k_i$ and $D_{ii}$ $\forall i=1,2,\dots,n$ given by $k_{l,i},k_{u,i}$ and $0,D_{\text{max}}$, respectively. Note $D_{ii}\geq0$ because the temperature cannot be negative. The values of $k_{l,i}$ and $k_{u,i}$ are determined such that the following stability condition is satisfied
\begin{equation}
    |s\mathbf{I}-\mathbf{A}_{cl}|\neq 0 \quad \forall s\in\mathbb{C} \quad \text{s.t.} \quad \Re{s}>0.
\end{equation}
Using $\omega=1,\gamma=2,\mu_1(0)=0.5,\mu_2(0)=0.7,\Sigma_{11}=\Sigma_{22}=0.01,\Sigma_{12}=\Sigma_{21}=0$, $D_{\text{max}}=\infty$, $k_{l,1}=-1,k_{l,2}=-2,k_{u,1}=k_{u,2}=\infty$ and
$u(t)=-[k_1 \quad k_2]\begin{bmatrix}
x_1\\x_2
\end{bmatrix}$ in system \eqref{bmotiona}, we have explored the solution of Equations \eqref{opt1} and \eqref{opt2} via the MATLAB Toolbox FMINCON \cite{matlabfmincon}. The solutions give us the set values of $\mathbf{k}$ and $\mathbf{D}$ that give at least a local minimum. Note that our goal here is to see the implications of a solution to such problems instead of rigorously finding the global optimal solution.
\begin{figure}[h]
    \centering
    \includegraphics[trim={1.5cm 1cm 2cm 0.5cm},clip,width=\columnwidth]{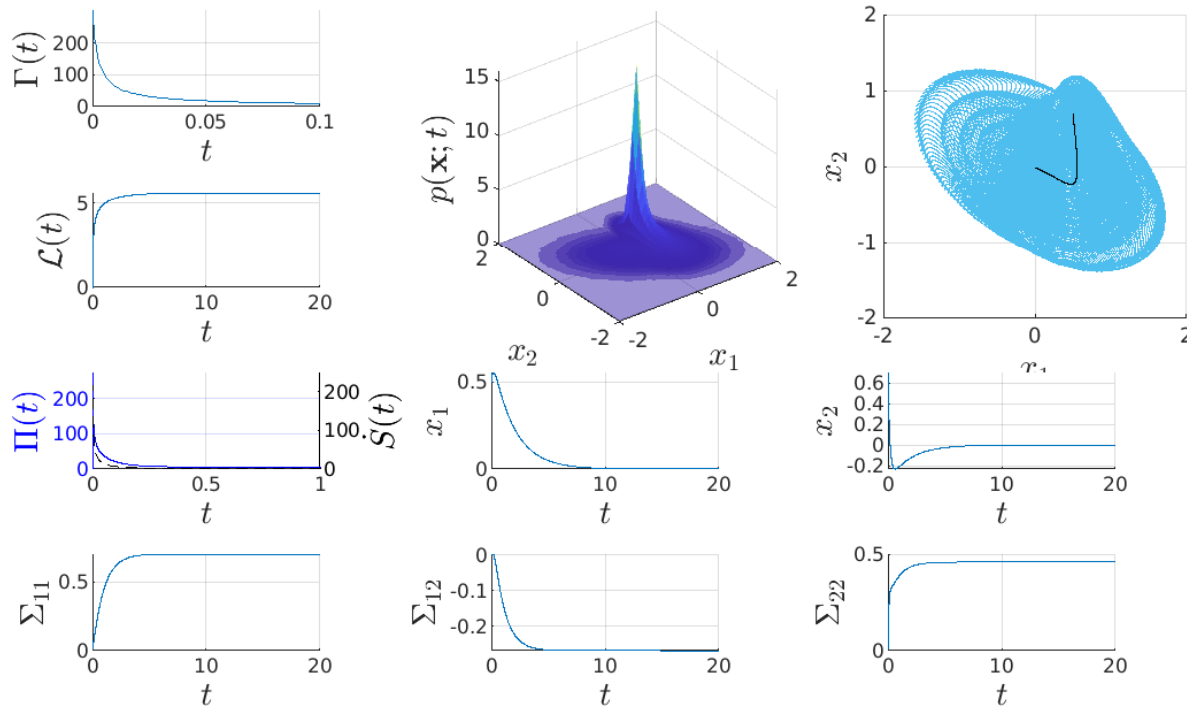}
    \caption{Full-state feedback control and temperature setting minimising $J_1$. A local minima is at $\mathbf{k}=[2.1229,4.4453]^\top$ and $\mathbf{D}=[0.2684,0;0,2.1181]$. The system parameters and initial conditions are $\omega=1,\gamma=2,\mu_1(0)=0.5,\mu_2(0)=0.7,\Sigma_{11}=\Sigma_{22}=0.01,\Sigma_{12}=\Sigma_{21}=0$.}
    \label{fig:opt1}
\end{figure}
Figure \ref{fig:opt1} depicts the time evolution of $\Gamma,\mathcal{L},\Pi,x_1,x_2$ and the spaces $(p(\mathbf{x},t),x_1,x_2)$ and $(x_1,x_2)$ after applying the values of $\mathbf{k}$ and $\mathbf{D}$ that give a solution to the optimisation problem \eqref{opt1}. As a result, the value of $\mathbf{k}$ contains $k_1=2.1229$ and $k_2=4.4453$ and $\mathbf{D}$, contains $D_{11}=0.2684$ and $D_{22}=2.1181$. This values produce an abrupt change in $\Gamma$, a quasi-logarithmic change in $\mathcal{L}$ with a maximum value slightly over $5$ and a slow almost critically damped change in the system dynamics towards the equilibrium. In addition, the control quickly drives $\Pi$ and $\dot{S}$ to zero. Even though the control action imposes a slow evolution of the mean value, the information rate quickly decreases. Such behaviour is desirable for systems where minimum information variability is more important than the speed under which we reach the equilibrium.

\begin{figure}[h]
    \centering
    \includegraphics[trim={1.5cm 1cm 1.8cm 0.5cm},clip,width=\columnwidth]{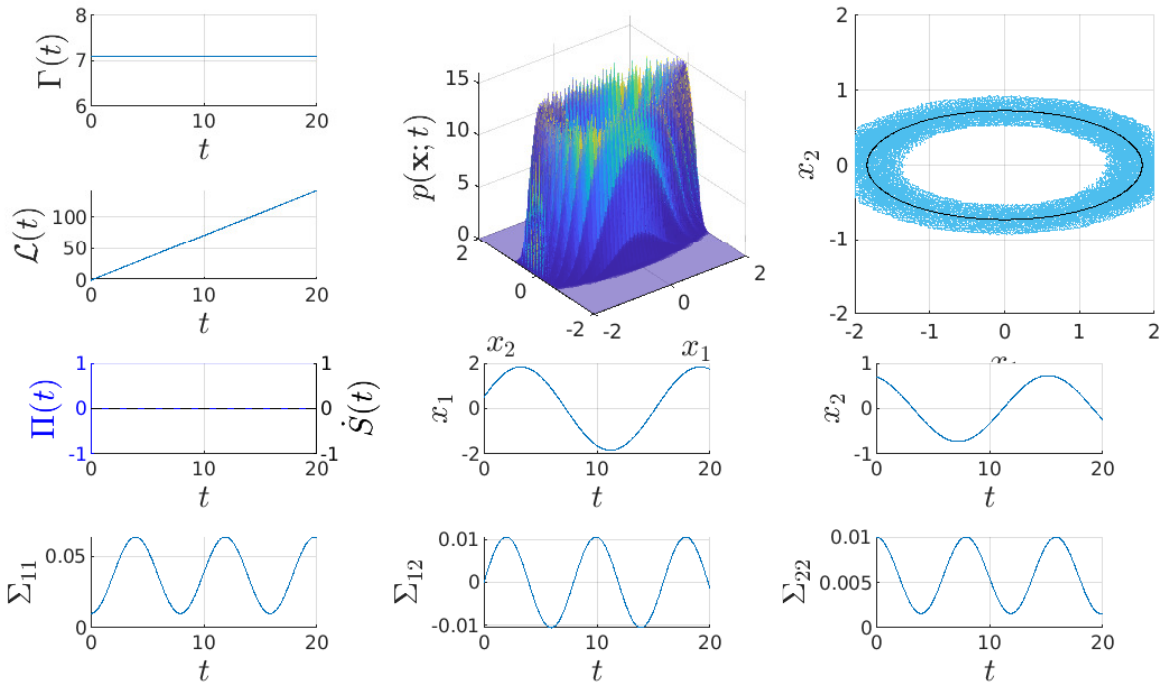}
    \caption{Full-state feedback control and temperature setting minimising $J_2$. A local minima is at $\mathbf{k}=[-0.8431 ,-2]^\top$ and $\mathbf{D}=[0,0;0,0]$. The system parameters and initial conditions are $\omega=1,\gamma=2,\mu_1(0)=0.5,\mu_2(0)=0.7,\Sigma_{11}=\Sigma_{22}=0.01,\Sigma_{12}=\Sigma_{21}=0$.}
    \label{fig:opt3}
\end{figure}

The solution of the optimisation problem \eqref{opt2} depicted in Figure \ref{fig:opt3} shows that a geodesic solution is obtained when entropy production $\Pi$ and entropy rate $\dot{S}$ are zero. This is imposed by the resultant control parameters $k_1=-0.8431$, $k_2=-2$, $D_{11}=0$ and $D_{22}=0$ which generate a harmonic oscillatory behaviour of the mean value $\boldsymbol{\mu}=[\mu_1,\mu_2]^\top$ and small changes in the time evolution of the covariance matrix elements $\Sigma_{11},\Sigma_{12}$ and $\Sigma_{22}$ to keep $\Gamma$ constant at all $t$. Note that, $D_{11}=D_{22}=0$ just means the absence of any external stochastic noise. However, there is stochasticity in the system due to the stochasticity in the initial condition (that is, our initial PDFs have a finite width). This optimal result is similar to the one shown in \cite{Proesmans2022} where authors find that the optimal energy landscape corresponds to an harmonic oscillator at all times.
\section{Conclusions}\label{sec:concl}
We have derived relations between information rate, entropy production and entropy rate for linear stochastic Gaussian processes changing according to the structure of the harmonic potential. Given such information-thermodynamic connection, we explore the applicability of these quantities to the design of thermodynamic optimal control algorithms and the detection of abrupt events in models of different nature. The results demonstrate that information rate $\rho_\Gamma$ and entropy production $\rho_\Pi$ correlation coefficients predict and detect abrupt events in the first and second moments of the stochastic dynamics, respectively. For higher-order systems, the norm of the information/thermodynamic quantities represents a fair approximation of the correlation between all the system random variables. Besides, the control applications show that it is possible to obtain the geodesic of the information length via a simple full-state feedback control algorithm. 

Since the proposed results permit us to establish a clear connection between information geometry and thermodynamic quantities, we can undoubtedly create cost functions that lead to energetically efficient (minimum entropy production) and organised (with minimum information variability) behaviours through IL. In the future, we plan to explore further connections with the area of control theory for applications in more complex scenarios. For instance, we can extend our results to non-linear systems by employing approximation methods such as the Laplacian assumption \cite{marreiros2009population,da2021bayesian,GuelCortez2022a}. Then, we can utilise modern control techniques such as the model-predictive-control \cite{camacho2013model,GuelCortez2022b} to find the solution to the proposed optimisation problems $J_1$ and $J_2$. This could bring potential benefits in the research areas of population dynamics \cite{schwartenbeck2013exploration} or inference control \cite{lanillos2021active,baltieri2019pid}.

\appendix
\section{Derivation of Entropy rate for Gaussian Process}\label{sec:ERProof}

We start by applying the definition of entropy \eqref{eprod} production and entropy flux \eqref{efluxb} giving us  
\begin{eqnarray}
\hspace{-1cm}\Pi_{J_i}\!\!&=&\!\!\frac{1}{D_{ii}}\left\langle f_i(\mathbf{x},t)^2\right\rangle\!+\!D_{ii}\left\langle\! \left(\frac{\partial Q(\mathbf{x})}{\partial x_i}\right)^2\right\rangle\!+\!2\left\langle \frac{\partial f_i(\mathbf{x},t)}{\partial x_i}\right\rangle, \label{piji}\\
\hspace{-1cm}\Phi_{J_i}\!\!&=&\!\!\frac{1}{D_{ii}}\langle f_i(\mathbf{x},t)^2\rangle+\left\langle \frac{\partial f_i(\mathbf{x},t)}{\partial x_i}\right\rangle.
\end{eqnarray}
Before continuing, it is useful to note that \cite{petersen2008matrix}
\begin{equation}
    \frac{\partial Q}{\partial x_k}=-\frac{1}{2}\left[\sum_i \delta{x}_i\Sigma_{ki}^{-1}+\sum_j \delta{x}_j\Sigma_{jk}^{-1}\right]=-\sum_i\delta x_i \Sigma_{ki}^{-1}=-\delta\mathbf{x}^\top\boldsymbol{\Sigma}_k^{-1}
\end{equation}
where $\delta{x}_i=x_i-\mu_i$, $\delta \mathbf{x}:=\mathbf{x}-\boldsymbol{\mu}=[\delta x_1,\dots,\delta x_n]^\top$ and $\boldsymbol{\Sigma}_k^{-1}$ is the $k$-th column of the inverse matrix $\boldsymbol{\Sigma}^{-1}$ of $\boldsymbol{\Sigma}$. Besides,
\begin{eqnarray}
f_i(\mathbf{x})^2\!\!=\!\!\mathbf{x}^\top\mathbf{A}_i^\top\mathbf{A}_i\mathbf{x}+\mathbf{u}^\top\mathbf{B}_i^\top\mathbf{B}_i\mathbf{u}+2\mathbf{u}^\top\mathbf{B}_i^\top\mathbf{A}_i\mathbf{x},
\end{eqnarray}
where we recall that $\mathbf{A}_i$ is the $i$-th arrow of the matrix $\mathbf{A}$.
Therefore \cite{petersen2008matrix}
\begin{equation}
    \left\langle D_{ii}\left(\frac{\partial Q(\mathbf{x})}{\partial x_i}\right)^2\right\rangle=D_{ii}\left\langle\delta\mathbf{x}^\top\boldsymbol{\Sigma}_i^{-1}(\boldsymbol{\Sigma}_i^{-1})^\top\delta\mathbf{x}\right\rangle =D_{ii}\Tr(\boldsymbol{\Delta}_i\boldsymbol{\Sigma}),
\end{equation}
and
\begin{equation}
    \frac{\left\langle f_i(\mathbf{x})^2\right\rangle}{D_{ii}}\!=\!\frac{1}{D_{ii}}\left(\Tr(\boldsymbol{\Gamma}_i\boldsymbol{\Sigma})+\boldsymbol{\mu}^\top\boldsymbol{\Gamma}_i\boldsymbol{\mu}+\mathbf{u}^\top\boldsymbol{\Omega}_i\mathbf{u}+2\mathbf{u}^\top\boldsymbol{\varphi}_i\boldsymbol{\mu}\right)
\end{equation}
where $\boldsymbol{\Delta}_i=\boldsymbol{\Sigma}_i^{-1}(\boldsymbol{\Sigma}_i^{-1})^\top$, $\boldsymbol{\Gamma}_i=\mathbf{A}_i^\top\mathbf{A}_i$, $\boldsymbol{\Omega}_i=\mathbf{B}_i^\top\mathbf{B}_i$, and $\boldsymbol{\varphi}_i=\mathbf{B}_i^\top\mathbf{A}_i$. 
Furthermore, we have that 
$$\frac{\partial f_i(\mathbf{x})}{\partial x_i}=a_{ii}.$$
Then
\begin{eqnarray}
\Pi_{J_i}\!&=&\!\frac{1}{D_{ii}}\left(\Tr(\boldsymbol{\Gamma}_i\boldsymbol{\Sigma})+\boldsymbol{\mu}^\top\boldsymbol{\Gamma}_i\boldsymbol{\mu}+\mathbf{u}^\top\boldsymbol{\Omega}_i\mathbf{u}+2\mathbf{u}^\top\boldsymbol{\varphi}_i\boldsymbol{\mu}\right)+D_{ii}\Tr(\boldsymbol{\Delta}_i\boldsymbol{\Sigma})\!\!+\!\!2a
_{ii}, \nonumber\\
\Phi_{J_i}\!&=&\!\frac{1}{D_{ii}}\left(\Tr(\boldsymbol{\Gamma}_i\boldsymbol{\Sigma})+\boldsymbol{\mu}^\top\boldsymbol{\Gamma}_i\boldsymbol{\mu}+\mathbf{u}^\top\boldsymbol{\Omega}_i\mathbf{u}+2\mathbf{u}^\top\boldsymbol{\varphi}_i\boldsymbol{\mu}\right)+a_{ii}.\label{Th1eqf}
\end{eqnarray}
Finally, since 
\begin{eqnarray}
\sum_i^n \frac{\boldsymbol{\mu}^\top\boldsymbol{\Gamma}_i\boldsymbol{\mu}}{D_{ii}} &=& \boldsymbol{\mu}^\top\left(\frac{\mathbf{A}_1^\top\mathbf{A}_1}{D_{11}}+\cdots+\frac{\mathbf{A}_n^\top\mathbf{A}_n}{D_{nn}}\right)\boldsymbol{\mu}=\boldsymbol{\mu}^\top\mathbf{A}^\top\mathbf{D}^{-1}\mathbf{A}\boldsymbol{\mu},\\
\sum_i^n \frac{\Tr(\boldsymbol{\Delta}_i\boldsymbol{\Sigma})}{D_{ii}}\!&=&\!\Tr\left(\left(\frac{\boldsymbol{\Sigma}_1^{-1}(\boldsymbol{\Sigma}_1^{-1})^\top}{D_{11}}+\cdots+\frac{\boldsymbol{\Sigma}_n^{-1}(\boldsymbol{\Sigma}_n^{-1})^\top}{D_{nn}}\right)\boldsymbol{\Sigma}\right)\nonumber\\
&=&\Tr\left(\boldsymbol{\Sigma}^{-1}\mathbf{D}^{-1}(\boldsymbol{\Sigma}^{-1})^\top\boldsymbol{\Sigma}\right).
\end{eqnarray}
Now, after applying the same reasoning to all the terms in the right hand side of \eqref{Th1eqf} and simplifying, we get
\begin{eqnarray}
\Pi&=&\dot{\boldsymbol{\mu}}^\top\mathbf{D}^{-1}\dot{\boldsymbol{\mu}}\!+\!\Tr\left(\mathbf{A}^\top\mathbf{D}^{-1}\mathbf{A}\boldsymbol{\Sigma}\right)+\Tr\left(\boldsymbol{\Sigma}^{-1}\mathbf{D}\right)\!+\!2\Tr(\mathbf{A}), \label{SPeqa}\\
\Phi&=&\dot{\boldsymbol{\mu}}^\top\mathbf{D}^{-1}\dot{\boldsymbol{\mu}}\!+\!\Tr\left(\mathbf{A}^\top\mathbf{D}^{-1}\mathbf{A}\boldsymbol{\Sigma}\right)
+\Tr(\mathbf{A}), \label{SFeqa}\\
\dot{S}&=&\Tr\left(\boldsymbol{\Sigma}^{-1}\mathbf{D}\right)\!+\!\Tr(\mathbf{A})=\frac{1}{2}\Tr\left(\boldsymbol{\Sigma}^{-1}\dot{\boldsymbol{\Sigma}}\right), \label{TSeqa}
\end{eqnarray}
which corresponds to the result given in Relation \ref{EPEFGP}. 

To provide extra confidence in the validity of expressions \eqref{SPeqa}-\eqref{TSeqa}. Consider as an example, the stochastic simulation \cite{Thiruthummal2022}, i.e. the multiple numerical solution of the stochastic differential equation, of \eqref{bmotiona} using the parameters $\boldsymbol{\mu}(0)=[1,0.5]^\top,a_{11}=0,a_{12}=1,a_{21}=-2,a_{22}=-3$, $\mathbf{D}=0.01\mathbf{I}$ and $\boldsymbol{\Sigma}(0)=0.1\mathbf{I}$ to compute the value of entropy production $\Pi$ via \eqref{piji}. Then, let us compare it with the ``analytical'' solution obtained using \eqref{SPeqa} and the solution of equations \eqref{eqmdynamics2}-\eqref{eqsdynamics2}. 
\begin{figure}[h!]
    \centering
    \begin{subfigure}[b]{0.7\columnwidth}
    \includegraphics[trim={1cm 0cm 1.5cm 0cm},clip,width=\columnwidth]{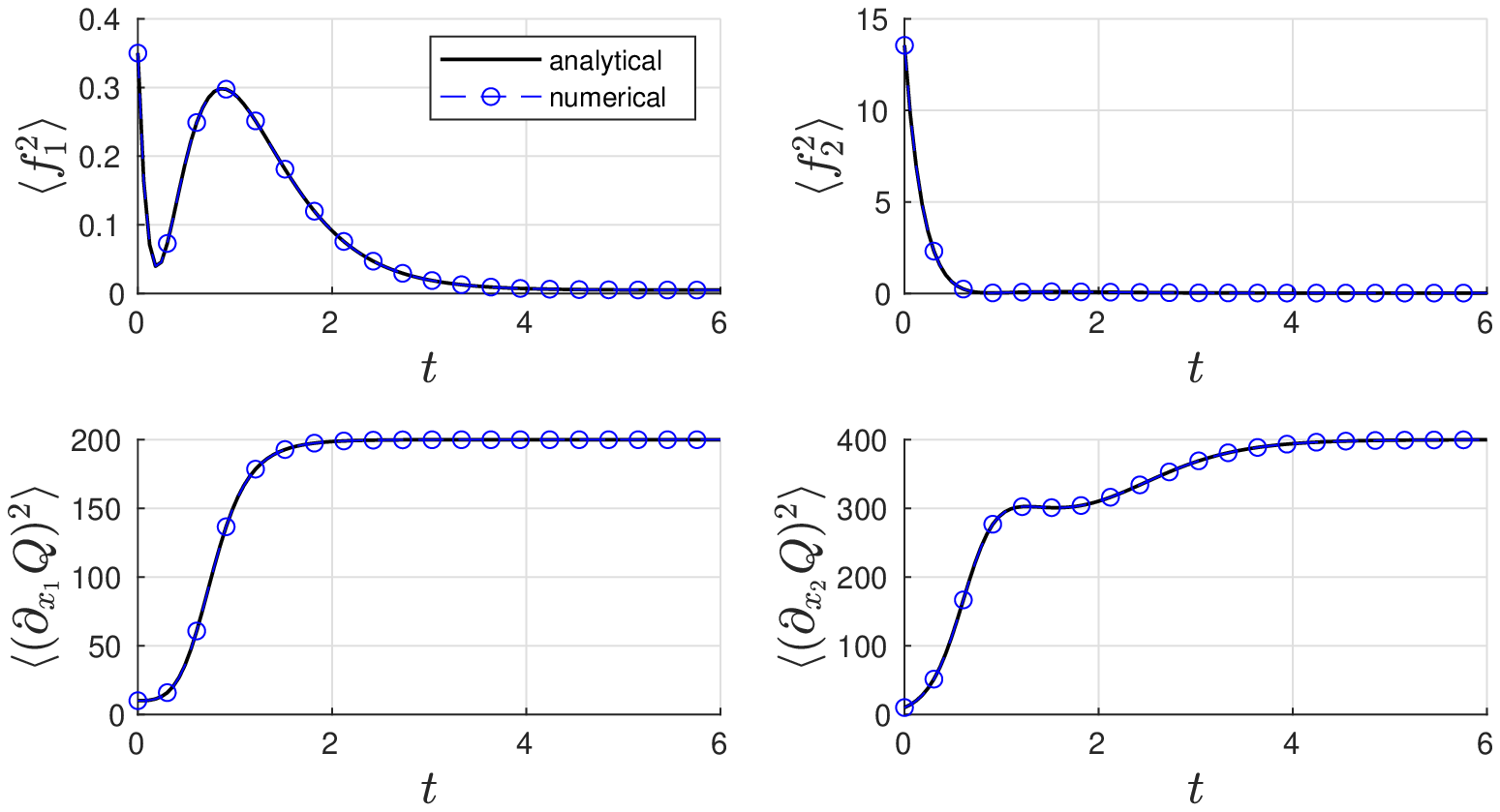}\vspace{-0.7cm}
     \caption{}
    \end{subfigure}
    \begin{subfigure}[b]{0.7\columnwidth}
    \includegraphics[trim={0.5cm 0cm 1.5cm 0cm},clip,width=\columnwidth]{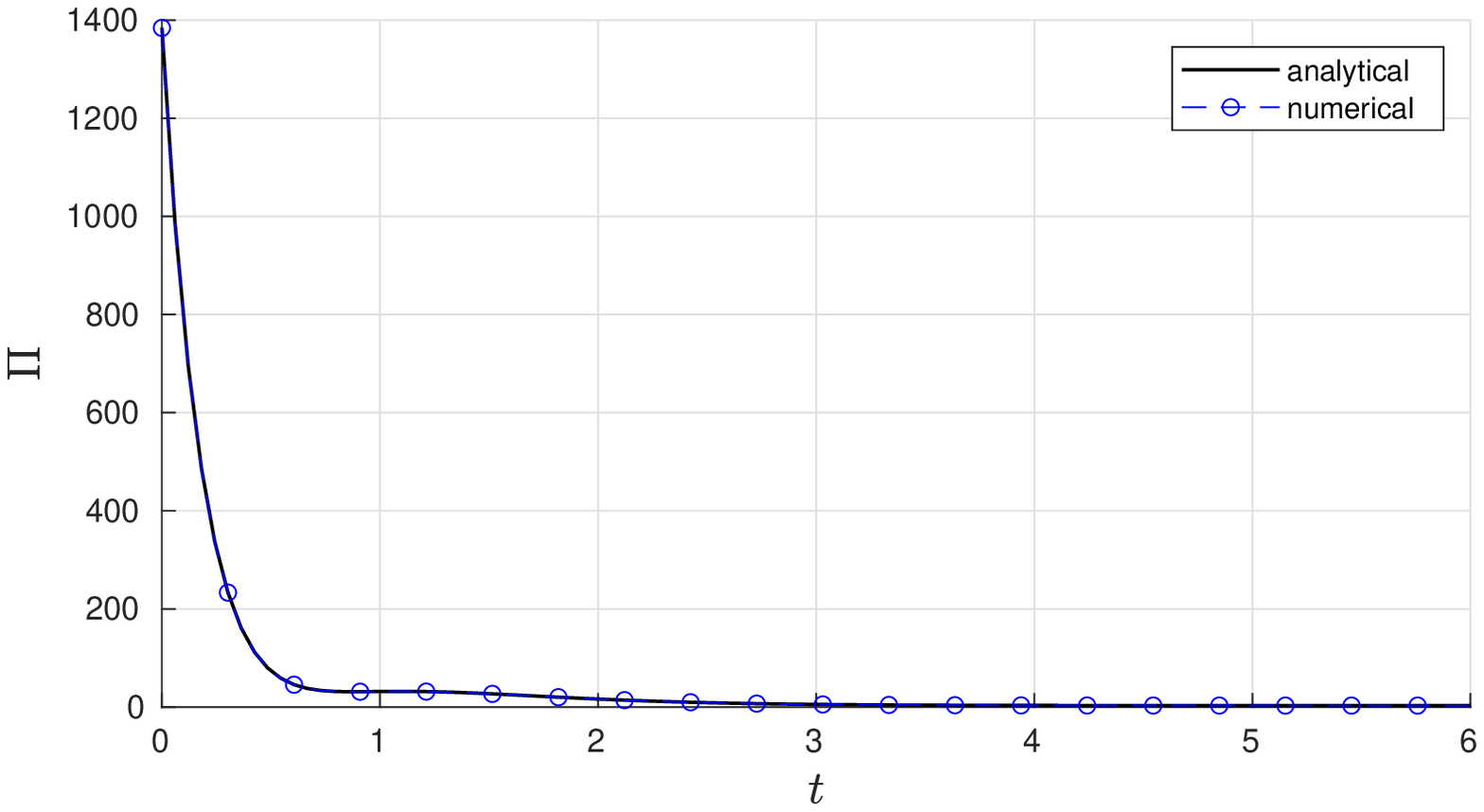} \vspace{-0.7cm}
     \caption{}
    \end{subfigure}    
    \caption{Stochastic simulation (numerical) vs analytical solution of the entropy production $\Pi$ for \eqref{bmotiona} using the parameters $\boldsymbol{\mu}(0)=[1,0.5]^\top,a_{11}=0,a_{12}=1,a_{21}=-2,a_{22}=-3$, $\mathbf{D}=0.01\mathbf{I}$ to produce $\boldsymbol{\Sigma}(0)=0.1\mathbf{I}$. The results show an agreement between the analytical and numerical results as expected. }
    \label{fig:appproof}
\end{figure}

Since the stochastic simulation uses its solutions to compute the averages $\langle f_i\rangle$ and $\langle(\partial_{x_i}Q)^2\rangle$ in \eqref{piji}. Figure \ref{fig:appproof} includes a comparison of the analytical vs numerical values of $\langle f_1\rangle, \langle f_2\rangle,\langle(\partial_{x_1}Q)^2\rangle, \langle(\partial_{x_2}Q)^2\rangle$ and $\Pi$ showing a perfect agreement between both approaches as expected.

\section{Proof of Relation \ref{mainth}}\label{sec:appproofmainth}
For any real matrix $\mathbf{A}$ in system \eqref{eqmdynamics2}-\eqref{eqsdynamics2}, we can rewrite the second term in the right hand side of \eqref{defirate} as follows 
\begin{eqnarray}
\Tr\left((\boldsymbol{\Sigma}^{-1}\dot{\boldsymbol{\Sigma}})^2\right)\!&=&\!\Tr\left(2\mathbf{A}^2+2\boldsymbol{\Sigma}^{-1}\mathbf{A}\boldsymbol{\Sigma}\mathbf{A}\right.\label{Epart2}\\
&&\hspace{-2.5cm}\left.+8\boldsymbol{\Sigma}^{-1}\mathbf{A}\mathbf{D}+4\boldsymbol{\Sigma}^{-1}\mathbf{D}\boldsymbol{\Sigma}^{-1}\mathbf{D}\right)\nonumber\\
&&\hspace{-2.5cm}=\Tr(2\mathbf{A}^2+4\boldsymbol{\Sigma}^{-1}\mathbf{A}\mathbf{D}+2\boldsymbol{\Sigma}^{-1}\mathbf{D}\boldsymbol{\Sigma}^{-1}\mathbf{D})\nonumber\\
&&\hspace{-2.5cm}+\Tr(2\boldsymbol{\Sigma}^{-1}\mathbf{A}\boldsymbol{\Sigma}\mathbf{A}+4\boldsymbol{\Sigma}^{-1}\mathbf{A}\mathbf{D}+2\boldsymbol{\Sigma}^{-1}\mathbf{D}\boldsymbol{\Sigma}^{-1}\mathbf{D})\nonumber\\
&&\hspace{-2.5cm}=2\Tr\left((\boldsymbol{\Sigma}^{-1}\mathbf{D}+A)^2\right) \nonumber\\
&&\hspace{-2.5cm}+2\Tr(\boldsymbol{\Sigma}^{-1}(\mathbf{A}\boldsymbol{\Sigma}\mathbf{A}^\top\mathbf{D}^{-1}\!+\!2\mathbf{A}\!+\!\mathbf{D}\boldsymbol{\Sigma}^{-1})\mathbf{D}).\nonumber
\end{eqnarray}
Equation \eqref{Epart2} can be written in terms of Entropy production $\Pi$ and entropy rate $\dot{S}$ using the following results.
First, from the fact that $\Pi\geq 0$ and $\Sigma^{-1},\mathbf{D}\succeq 0$ we get
\begin{eqnarray}
   \Tr(\boldsymbol{\Sigma}^{-1})\Pi\Tr(\mathbf{D})&=&\Tr(\boldsymbol{\Sigma}^{-1})\Tr\left(\dot{\boldsymbol{\mu}}\dot{\boldsymbol{\mu}}^\top\mathbf{D}^{-1}+\mathbf{A}\boldsymbol{\Sigma}\mathbf{A}^\top\mathbf{D}^{-1}+2\mathbf{A}+\mathbf{D}\boldsymbol{\Sigma}^{-1}\right)\Tr(\mathbf{D})\nonumber\\
    &&\hspace{-3cm}\geq\Tr(\boldsymbol{\Sigma}^{-1}\dot{\boldsymbol{\mu}}\dot{\boldsymbol{\mu}}^\top)+\Tr(\boldsymbol{\Sigma}^{-1}(\mathbf{A}\boldsymbol{\Sigma}\mathbf{A}^\top\mathbf{D}^{-1}\!+\!2\mathbf{A}\!+\!\mathbf{D}\boldsymbol{\Sigma}^{-1})\mathbf{D}).\label{Epart3}
\end{eqnarray}
Now,  taking $\lambda_i$ as the eigenvalues of the matrix $\mathbf{H}:=\boldsymbol{\Sigma}^{-1}\mathbf{D}+\mathbf{A}$, we have 
\begin{multline}
    \dot{S}^2=\Tr(\mathbf{H})^2=(\sum_{i=1}^n\lambda_i)^2=\sum_{i=1}^n\lambda_i^2+2\sum_{j=1}^n\sum_{i=1}^{j-1}\lambda_i\lambda_j
    =\Tr(\mathbf{H}^2)+2\sum_{i<j}^n\lambda_i\lambda_j=\Tr(\mathbf{H}^2)+2g(\mathbf{H}).\label{Epart4}
\end{multline}
Finally, using \eqref{Epart2}-\eqref{Epart4} in \eqref{defirate} we get
\begin{multline}
    \hspace{-0.4cm}2g(\mathbf{H})\!\leq\!\Gamma^2+2g(\mathbf{H})\!\leq\!\Tr(\boldsymbol{\Sigma}^{-1})\Pi\Tr(\mathbf{D})+\Tr(\mathbf{H}^2)+2g(\mathbf{H}), \\
    2g(\mathbf{H})\!\leq\!\Gamma^2+2g(\mathbf{H})\!\leq\!\Tr(\boldsymbol{\Sigma}^{-1})\Pi\Tr(\mathbf{D})+\dot{S}^2,\\
    0\!\leq\!\Gamma^2\!\leq\! \Tr(\boldsymbol{\Sigma}^{-1})\Pi\Tr(\mathbf{D})+\dot{S}^2-2g(\mathbf{H}).
\end{multline}
Now, since $\dot{S}=\sum_i^n\dot{S}_{J_i}$ where again $\dot{S}_{J_i}$ is the contribution of the current flow $J_i$ to the total entropy rate $\dot{S}$, we see that each eigenvalue $\lambda_i=\dot{S}_{J_i}$. Note that, for simplicity, we have changed the argument of the function $g$ from $\mathbf{H}$ to a new variable vector $\mathbf{s}=[\dot{S}_{J_1},\dot{S}_{J_2},\dots,\dot{S}_{J_n}]^\top$, i.e. now $g:\mathbf{R}^n\to\mathbf{R}$. This ends our proof.

\section{Proof Relation \ref{mainth2}}\label{sec:IneProof}
To derive the result shown in Relation \ref{mainth}, we first consider the following preliminary results \cite{yang2000matrix,shebrawi2013trace,patel1979trace}
\begin{eqnarray}
    \Tr(\mathbf{X}\mathbf{Y})\!&\leq&\!\Tr(\mathbf{X})\Tr(\mathbf{Y}) \quad\forall \quad\mathbf{X},\mathbf{Y}\succeq 0,\\
    \Tr(\dot{\boldsymbol{\Sigma}})\!&=&\!\Tr(\mathbf{A}\boldsymbol{\Sigma})\!+\!\Tr(\boldsymbol{\Sigma}\mathbf{A}^\top)+\Tr(2\mathbf{D})\nonumber\\
    \!&=&\!2\Tr(\boldsymbol{\Sigma}\mathbf{A}+\mathbf{D}),\\
    \Tr(\dot{\boldsymbol{\Sigma}})^2\!&=&\!4\Tr(\boldsymbol{\Sigma}\mathbf{A}+\mathbf{D})\Tr(\boldsymbol{\Sigma}\mathbf{A}^\top+\mathbf{D})\nonumber\\
    \!&\geq&\!4 \Tr(\boldsymbol{\Sigma}\mathbf{A}\boldsymbol{\Sigma}\mathbf{A}^\top\!+\!2\boldsymbol{\Sigma}\mathbf{A}\mathbf{D}\!+\!\mathbf{D}^2),\\
    \Tr(\boldsymbol{\Sigma}^{-1})^2\Tr(\dot{\boldsymbol{\Sigma}})^2\!&=&\!\Tr(\boldsymbol{\Sigma}^{-1})\Tr(\dot{\boldsymbol{\Sigma}})\Tr(\boldsymbol{\Sigma}^{-1})\Tr(\dot{\boldsymbol{\Sigma}})\nonumber\\
    \!&\geq&\!\Tr(\boldsymbol{\Sigma}^{-1}\dot{\boldsymbol{\Sigma}})^2\geq \Tr((\boldsymbol{\Sigma}^{-1}\dot{\boldsymbol{\Sigma}})^2)
\end{eqnarray}

Then, by applying the previous results to the definition of $\Gamma^2$ in \eqref{defirate}, we have
\begin{eqnarray}
0\leq\Gamma^2\!&=&\!\Tr(\dot{\boldsymbol{\mu}}^\top\boldsymbol{\Sigma}^{-1}\dot{\boldsymbol{\mu}})+\frac{1}{2}\Tr\left((\boldsymbol{\Sigma}^{-1}\dot{\boldsymbol{\Sigma}})^2\right) \nonumber \\
\!&\leq&\!\Tr(\dot{\boldsymbol{\mu}}\dot{\boldsymbol{\mu}}^\top)\Tr(\boldsymbol{\Sigma}^{-1})+\frac{1}{2}\Tr(\boldsymbol{\Sigma}^{-1}\dot{\boldsymbol{\Sigma}})^2\label{ineproof2a} \\
\!&\leq&\!\Tr(\dot{\boldsymbol{\mu}}\dot{\boldsymbol{\mu}}^\top)\Tr(\boldsymbol{\Sigma}^{-1})+\frac{1}{4}\Tr(\boldsymbol{\Sigma}^{-1})^2\Tr(\dot{\boldsymbol{\Sigma}})^2+\dot{S}^2.\nonumber
\end{eqnarray}
Now, multiplying both sides of inequality \eqref{ineproof2a} by $\Tr(\mathbf{D}^{-1}\mathbf{D})$ and factorising $\Tr(\boldsymbol{\Sigma}^{-1})$ from its right hand side, we have
\begin{eqnarray}
0\leq n\Gamma^2\!&\leq&\!\Tr(\boldsymbol{\Sigma}^{-1})\{\Tr(\dot{\boldsymbol{\mu}}\dot{\boldsymbol{\mu}}^\top)\Tr(\mathbf{D}^{-1}\mathbf{D})\nonumber\\
\!&&\!+\frac{1}{4}\Tr(\boldsymbol{\Sigma}^{-1})\Tr(\dot{\boldsymbol{\Sigma}})^2\Tr(\mathbf{D}^{-1}\mathbf{D})\}+n\dot{S}^2\nonumber \\
\!&\leq&\!\Tr(\boldsymbol{\Sigma}^{-1})\{\Tr(\dot{\boldsymbol{\mu}}\dot{\boldsymbol{\mu}}^\top)\Tr(\mathbf{D}^{-1})\label{ineproof2b}\\
\!&&\!+\frac{1}{4}\Tr(\boldsymbol{\Sigma}^{-1})\Tr(\dot{\boldsymbol{\Sigma}})^2\Tr(\mathbf{D}^{-1})\}\Tr(\mathbf{D})+n\dot{S}^2.\nonumber
\end{eqnarray}
From the right hand side of \eqref{ineproof2b}, we define the part inside the curly brackets as
\begin{equation}
\Pi_u\!:=\!\Tr(\dot{\boldsymbol{\mu}}\dot{\boldsymbol{\mu}}^\top)\Tr(\mathbf{D}^{-1})+\frac{1}{4}\Tr(\boldsymbol{\Sigma}^{-1})\Tr(\dot{\boldsymbol{\Sigma}})^2\Tr(\mathbf{D}^{-1}). \label{ineproof2c}
\end{equation}
Which gives us the expression in our result \eqref{mainresult}. The value of $\Pi_u$ can be proved to be an upper bound of $\Pi$ from the following reasoning
\begin{eqnarray}
\Pi_u&\geq&\Tr(\dot{\boldsymbol{\mu}}^\top\mathbf{D}^{-1}\dot{\boldsymbol{\mu}})\nonumber\\
&&\hspace{-0.4cm}+\Tr(\boldsymbol{\Sigma}^{-1})\Tr(\boldsymbol{\Sigma}\mathbf{A}\boldsymbol{\Sigma}\mathbf{A}^\top\!+\!2\boldsymbol{\Sigma}\mathbf{A}\mathbf{D}\!+\!\mathbf{D}^2)\Tr(\mathbf{D}^{-1}) \nonumber\\
&&\hspace{-0.7cm}\geq\dot{\boldsymbol{\mu}}^\top\mathbf{D}^{-1}\dot{\boldsymbol{\mu}}\!+\!\mathbf{tr}\left(\boldsymbol{\Sigma}^{-1}\boldsymbol{\Sigma}(\mathbf{A}\boldsymbol{\Sigma}\mathbf{A}^\top\mathbf{D}^{-1}\!+\!2\mathbf{A}\!+\!\boldsymbol{\Sigma}^{-1}\mathbf{D})\mathbf{D}\mathbf{D}^{-1}\right)\nonumber\\
&&\hspace{-0.7cm}=\dot{\boldsymbol{\mu}}^\top\mathbf{D}^{-1}\dot{\boldsymbol{\mu}}+\mathbf{tr}(\mathbf{A}\boldsymbol{\Sigma}\mathbf{A}^\top\mathbf{D}^{-1}+2\mathbf{A}+\boldsymbol{\Sigma}^{-1}\mathbf{D}). \label{ineproof2d}
\end{eqnarray}
Note that for $\Pi_u\geq\Pi$ we need $\mathbf{A}\succeq 0$.
A similar result can be found starting from the definition of $\Pi_u$ in \eqref{ineproof2c} as follows
\begin{multline}
\Tr(\boldsymbol{\Sigma}^{-1})\Pi_u\Tr(\mathbf{D})\!\geq\!\Tr(\boldsymbol{\Sigma}^{-1})\{\Tr(\dot{\boldsymbol{\mu}}\dot{\boldsymbol{\mu}}^\top)\Tr(\mathbf{D}^{-1}\mathbf{D})\\
+\frac{1}{4}\Tr(\boldsymbol{\Sigma}^{-1})\Tr(\dot{\boldsymbol{\Sigma}})^2\Tr(\mathbf{D}^{-1}\mathbf{D})\}. \label{ineproof2e}
\end{multline}
From \eqref{ineproof2d}, the main result follows straightforwardly using \eqref{ineproof2b} leading to our main result in Relation \ref{mainth}.
\section{Proof of Relation \ref{th3}}\label{sec:EquProof}
If $\mathbf{A}$ is an $n\times n$ diagonal matrix, then $\boldsymbol{\Sigma}$ and $\dot{\boldsymbol{\Sigma}}$ are also diagonal and the following expressions hold
\begin{eqnarray}
\Gamma^2&=&\sum_i\Gamma_{i}^2=\sum_{i}\left(\frac{\dot{\mu}_i^2}{\Sigma_{ii}}+\frac{1}{2}\left(\frac{\dot{\Sigma}_{ii}}{\Sigma_{ii}}\right)^2\right), \label{eqinde}\\
\Pi&=&\sum_i\Pi_{i}=\dot{\boldsymbol{\mu}}^\top\mathbf{D}^{-1}\dot{\boldsymbol{\mu}}+\frac{1}{4}\Tr\left(\boldsymbol{\Sigma}^{-1}\dot{\boldsymbol{\Sigma}}^2\mathbf{D}^{-1}\right)\nonumber\\
&=&\sum_i\left(\frac{\dot{\mu}_i^2}{D_{ii}}+\frac{\dot{\Sigma}_{ii}^2}{4\Sigma_{ii}D_{ii}}\right),\label{eqindpi}\\
\dot{S}&=&\sum_i\dot{S}_{i}=\frac{1}{2}\sum_i\frac{\dot{\Sigma}_{ii}}{\Sigma_{ii}}. \label{eqindS}
\end{eqnarray}
By rearranging Equations \eqref{eqindpi} and \eqref{eqindS} to form $\Gamma^2$, we have
\begin{equation}
    \Gamma_{i}^2=\frac{D_{ii}}{\Sigma_{ii}}\Pi_{i}+\dot{S}_{i}^2.
\end{equation}
which leads to our result.

\section{Proof of equation \eqref{pitf}}\label{sec:Equlimit}
Given that $\mathbf{A}$ is a diagonal and stable matrix, we have
\begin{eqnarray}
\dot{\boldsymbol{\Sigma}}&=&2\mathbf{A}\boldsymbol{\Sigma}+2\mathbf{D},\\
\dot{\boldsymbol{\mu}}&=&\mathbf{A}\boldsymbol{\mu},
\end{eqnarray}
which gives
\begin{eqnarray}
\boldsymbol{\Sigma}^{-1}(\infty)&=&-\mathbf{D}^{-1}\mathbf{A},\\
\dot{\boldsymbol{\Sigma}}(\infty)&=&0,\\
\boldsymbol{\mu}(\infty)&=&0,\\
\dot{\boldsymbol{\mu}}(\infty)&=&0.
\end{eqnarray}
In this manner, the entropy production and entropy rate are
\begin{eqnarray}
\Pi(\infty)&=&\sum_i\left(\frac{\dot{\mu}_i(\infty)^2}{D_{ii}}+\frac{\dot{\Sigma}_{ii}(\infty)^2}{4\Sigma_{ii}(\infty)D_{ii}}\right)=0,\\
\dot{S}(\infty)&=&\sum_i\dot{S}_{J_i}=\frac{1}{2}\sum_i\frac{\dot{\Sigma}_{ii}(\infty)}{\Sigma_{ii}(\infty)}=0.
\end{eqnarray}
Giving us the result mentioned in equation \eqref{pitf}.
\begin{equation}
    \boldsymbol{\Sigma}^{-1}(\infty)=-\mathbf{D}^{-1}\mathbf{A}, \quad\Pi(\infty)=0.\quad \mathbf{s}(\infty)=0. 
\end{equation}
\end{document}